\documentclass[journal,10pt]{IEEEtran}
\IEEEoverridecommandlockouts
\ifCLASSINFOpdf
\else
\fi
\usepackage{amsmath}
\usepackage{cite}
\usepackage{algorithm}
\usepackage[noend]{algpseudocode}
\usepackage{graphicx}
\usepackage{textcomp}
\usepackage{xcolor}
\usepackage[nolist,nohyperlinks]{acronym}
\usepackage{lipsum}  
\usepackage{float}
\usepackage{bm}
\usepackage{subfigure} 
\usepackage{flushend}
\usepackage{cuted}
\usepackage{wrapfig}
\usepackage{needspace}
\usepackage{rotating}
\usepackage{bbm}
\usepackage[utf8x]{inputenc}
\usepackage{tikz}
\usepackage{enumerate}
\usepackage[inline]{enumitem}
\usepackage{multirow}
\usepackage{array}
\usepackage{colortbl}
\usepackage{verbatim}
\usepackage{url}
\usepackage{soul}
\definecolor{LightBlue}{rgb}{0.67,0.847,0.9}
\definecolor{LightOrange}{rgb}{1,0.84,0.60}
\definecolor{LightGrey}{rgb}{0.9,0.9,0.9}

\newcommand{\ti}[1]{\textit{#1}}
\newcommand{\mq}[1]{$`{#1}$'}

\newcommand{\bracket}[1]{\left({#1}\right)}

\newcommand{\argmin}[1]{\underset{#1}{\operatorname{arg}\,\operatorname{min}}\;}

\begin{document}
	
	%\title{A Mobility Adapted Smart ABR Video Streaming Framework for mmWave Networks}
	%\title{Network-Application Integration over 5G mmWave for Energy-Efficient Video Streaming}
	\title{Improving UE Energy Efficiency through Network-aware Video Streaming over 5G}
	% {\footnotesize \textsuperscript{*}%Note: Sub-titles are not captured in Xplore and
	% %should not be used
	% }
	
\author{Basabdatta~Palit,
        Argha Sen,
		Abhijit~Mondal,~\IEEEmembership{Member,~IEEE,}
		Ayan~Zunaid,
		Jay~Jayatheerthan,
		and~Sandip Chakraborty,~\IEEEmembership{Member,~IEEE}
	\thanks{\noindent Basabdatta Palit is with the Department of Electronics and Telecommunication Engineering, Indian Institute of Engineering, Science and Technology, Shibpur, Howrah 711103, India. Email: basabdatta.iitkgp@gmail.com}
	\thanks{\noindent Argha Sen, Ayan Zunaid, and Sandip Chakraborty are with the Department of Computer Science and Engineering, Indian Institute of Technology, Kharagpur, West Bengal, India. Email: arghasen10@gmail.com, ayanzunaid10@gmail.com, sandipc@cse.iitkgp.ac.in}
	\thanks{\noindent Abhijit Mondal is with VDX.tv. Email: am@abhijitmondal.in}
	\thanks{\noindent Jay Jayatheerthan,  is with Intel Technology Pvt. Ltd. India, E-mail: jay.jayatheerthan@intel.com}% <-this % stops a space
}
%\author{Argha Sen$^1$,
%	Abhijit Mondal$^2$,~\IEEEmembership{Member,~IEEE,}
%	Basabdatta Palit$^3$,~\IEEEmembership{Member,~IEEE,}
%	Jay Jayatheerthan$^4$, Krishna Paul$^5$
%	and~Sandip Chakraborty$^6$,~\IEEEmembership{Member,~IEEE}% <-this % stops a space
%	\thanks{ Argha Sen, Abhijit Mondal, Sandip Chakraborty are with the Department
%		of Computer Science and Engineering, Indian Institute of Technology, Kharagpur,
%		Kharagpur - 721302, India. Basabdatta Palit is with the Department
%		of Electronics and Telecommunication Engineering, Indian Institute of Engineering, Science and Technology, Shibpur, Howrah 711103, India. Jay Jayatheerthan,  Krishna Paul are with Intel Technology Pvt. Ltd. India\\ e-mail: $^1$arghasen10@gmail.com, $^2$am@abhijitmondal.in, $^3$basabdatta.iitkgp@gmail.com, $^6$sandipc@cse.iitkgp.ac.in}% <-this % stops a space
%}
\maketitle

    \begin{abstract}
    Adaptive Bitrate (ABR) Streaming over the cellular networks has been well studied in the literature; however, existing ABR algorithms primarily focus on improving the end-users' Quality of Experience (QoE) while ignoring the resource consumption aspect of the underlying device. Consequently, proactive attempts to download video data to maintain the user's QoE often impact the battery life of the underlying device unless the download attempts are synchronized with the network's channel condition. In this work, we develop EnDASH-5G -- a wrapper over the popular DASH-based ABR streaming algorithm, which establishes this synchronization by utilizing a network-aware video data download mechanism. EnDASH-5G utilizes a novel throughput prediction mechanism for 5G mmWave networks by upgrading the existing throughput prediction models with a transfer learning-based approach leveraging publicly available 5G datasets. It then exploits deep reinforcement learning to dynamically decide the playback buffer length and the video bitrate using the predicted throughput. This ensures that the data download attempts get synchronized with the underlying network condition, thus saving the device's battery power. From a thorough evaluation of EnDASH-5G, we observe that it achieves a near $30.5\%$ decrease in the maximum energy consumption than the state-of-the-art Pensieve ABR algorithm while performing almost at par in QoE.
    
% 		User experience of watching videos in smartphones while travelling is often limited by fast battery drainage. Existing client video players use adaptive bitrate (ABR) streaming through Dynamic Adaptive Streaming over HTTP (DASH) to improve user's Quality of Experience (QoE) while ignoring the energy savings aspect, which has been addressed in our work. In this paper, we propose EnDASH-5G - an energy aware wrapper over DASH which minimizes energy consumption without compromising on QoE of users, under mobility. First, we undertake an extensive measurement study using two phones and three service providers to understand the dynamics between energy consumption of smartphones and radio related network parameters. Equipped with this study, the proposed system predicts cellular network throughput from the radio parameters within a finite future time window.  The prediction engine captures the effect of associated technology and vertical handovers on throughput, unlike existing works. EnDASH-5G then uses  deep reinforcement learning based neural networks to first tune the playback buffer length to the average predicted cellular network throughput and then to select an optimal video chunk bitrate. It achieves a near 27\% decrease in the maximum energy consumption than  state-of-the-art ABR Pensieve algorithm while performing almost at par in QoE.
    \end{abstract}

\begin{acronym}
	\acro{2G}{2$^\mathrm{nd}$ Generation}
	\acro{3G}{3$^\mathrm{rd}$ Generation}
	\acro{4G}{4$^\mathrm{th}$ Generation}
	\acro{5G}{5$^\mathrm{th}$ Generation}
	\acro{A3C}{Actor-Critic}
	\acro{ABR}{adaptive bitrate}
	\acro{BS}{Base Station}
	\acro{CDN}{Content Distribution Network}
	\acro{DASH}{Dynamic Adaptive Streaming over HTTP}
	\acro{DRX}{Discontinuous Reception}
	\acro{EDGE}{Enhanced Data Rates for \ac{GSM} Evolution.}
	\acro{FMPC}{Fast MPC}
	\acro{RMPC}{Robust MPC}
	
	\acro{eNB}{evolved NodeB}
	\acro{GSM}{Global System for Mobile}
	\acro{HD}{High Definition}
	\acro{HSPA}{High Speed Packet Access}
	\acro{LTE}{Long Term Evolution}	
	\acro{LSTM}{Long Short Term Memory}
	\acro{NR}{New Radio}
	\acro{HVPM}{High voltage Power Monitor}
	\acro{gNB}{gNodeB}
	\acro{QoS}{Quality of Service}
	\acro{QoE}{Quality of Experience}
	\acro{RF}{Random Forest}
	\acro{RFL}{Random Forest Learning}
	\acro{RL}{Reinforcement Learning}
	\acro{RRC}{Radio Resource Control}
	\acro{RSSI}{Received Signal Strength Indicator}
	\acro{RSRP}{Reference Signal Received Power}
	\acro{RSRQ}{Reference Signal Received Quality}
	\acro{SINR}{signal-to-interference-plus-noise-ratio}
	\acro{SNR}{signal-to-noise-ratio}
	\acro{UE}{User Equipment}
	\acro{UHD}{Ultra HD}
	\acro{VoLTE}{Voice over LTE}
	\acro{WiFi}{Wireless Fidelity}
	\acro{SUMO}{Simulation of Urban MObility}
	\acro{MCS}{Modulation Coding Scheme}
\end{acronym}

	\section{\textbf{Introduction}}\label{section:intro}
Sustenance in the new normal way of life stimulated by the global pandemic has been made possible by the pervasive usage of online video streaming services. These services range from official meetings and classes to the increase in screen time over \textit{Over the Top} (OTT) media services, such as Netflix, Amazon Prime, etc~\cite{marcos2020path}. The quality of the streaming sessions depends on the underlying network quality. Mobile Internet uses cellular connectivity, which may display unwarranted fluctuations and may have unidentified blind spots with little to no reception. In such cases, the users experience a very low throughput, which affects the quality of the streaming sessions. As cellular networks prepare to migrate to mmWave communication in the \ac{5G}, fluctuations are going to become more incessant \cite{Liu2020} due to the increased susceptibility of the high-frequency communication to path loss and scattering. 

Such fluctuations in the channel quality not only impact the \ac{QoE} for the video streaming but also affect the devices' energy consumption. Existing \ac{ABR} video streaming protocols are generously biased towards user experienced \ac{QoE} and minimizing data wastage than saving device energy consumption. To minimize the data wastage during the playback, the existing video players typically use \textit{fixed-size buffers} to download and store video chunks. The download of future video chunks is synchronized with the video playback -- the download is stalled until the user finishes watching a portion of the already downloaded video chunks. These approaches assume that the user may skip watching a part of the video, and, thus, such schemes can reduce data wastage. However, these impose an additional constraint that the next download session should initiate immediately after the available data in the playback buffer is depleted beyond a threshold. Consequently, the player is bound to initiate a download request even if the underlying network's signal quality is poor, thereby resulting in higher energy consumption at the \acp{UE}. {A recent measurement study on 5G device energy consumption~\cite{Narayanan2021} shows that services such as \ac{ABR} video streaming are more susceptible to increased energy usage in 5G mmWave networks.} It is, therefore, imperative to design energy-efficient video streaming algorithms for 5G mmWave networks~\cite{liu2021accelerating, yue2020energy, chen2020context}, especially to cater to the ubiquitous \ac{QoE} demand of vehicular \acp{UE}.

Supporting \ac{QoE} for video streaming and device's energy efficiency simultaneously needs a strong \textit{cross-layer} integration between the streaming application and the underlying network, as the download sessions need to be synchronized with the network quality. Existing streaming applications predominantly use \ac{DASH}~\cite{Stockhammer2011}, an \ac{ABR} streaming mechanism over HTTP. In \ac{DASH}, a target video is broken into chunks of fixed duration and stored at a \ac{CDN} server at different qualities. During a video session, the \ac{DASH} video player at the end-user device uses the estimated network throughput and the application layer playback buffer length to decide the quality (bitrate) at which the next chunk is to be fetched. However, to ensure that the device's energy efficiency improves, we argue that not only the bitrate for the next chunk but also the maximum playback buffer length needs to be tuned adaptively to the \textit{predicted network condition} for the near future. Thus, in line with our earlier work in~\cite{Mondal2020}, we propose \textit{EnDASH-5G}, a client-side energy-efficient video streaming algorithm that acts as a wrapper over \ac{DASH} -- custom-made for 5G mmWave communication. EnDASH-5G is essentially a rate adaptation mechanism, which pursues  opportune fetching of video chunks, tuned to the wireless link condition through the following operations -- (i) It first predicts the cellular throughput from radio-related parameters. (ii) It then uses the predicted throughput to estimate the maximum playback buffer length such that the energy usage is minimized. (iii) Finally, it chooses the optimal bitrate for the next video chunk to be downloaded using the estimated playback buffer length.
The earlier version EnDASH in~\cite{Mondal2020} was designed to suit the network quality variations of \ac{4G} only. The corresponding throughput prediction, as well as the buffer-length and bitrate prediction, are not directly extensible to \ac{5G} mmWave networks. The reasons are as follows.
\begin{enumerate}
    \item The throughput prediction mechanism used in EnDASH depends on 4G network parameters, such as received signal strength, base station characteristics, handover history, etc. These parameters are drastically different in a typical \ac{5G} network. In addition, 5G being a highly directional communication technology due to its high frequency of operation, other factors such as the number of antennas, the direction of antenna beams, etc., also impact the received throughput~\cite{Narayanan2021}. This inevitably implies that the throughput prediction model will need a complete redesigning. Furthermore, many of the input parameters to the throughput prediction engine of EnDASH are specific to a service provider or locality of operation. As the availability of large-scale commercial deployments of \ac{5G} networks is still scarce, it is difficult to have sufficient data for training a supervised learning model. This also calls for a new throughput prediction engine.
    \item The cross-layer integration of the video streaming at the application layer with the network needs a complete redesign as the network connectivity model is different in \ac{5G} compared to a 4G network. For example, \ac{5G} uses a more dense deployment of \acp{gNB} compared to \ac{4G} \acp{eNB}. Such an underlying deployment directly impacts the communication environment that needs to be considered during model development.
    \item The 5G NR \ac{UE} energy consumption model is also significantly different from \ac{4G}, which impacts the analysis and measurement of the device energy consumption during data communication. Hence, the model for energy analysis needs to be upgraded considering the \ac{5G} \ac{UE} energy states. 
\end{enumerate} 

Considering these factors, in this paper, we modify our earlier model in~\cite{Mondal2020} to EnDASH-5G -- an energy-efficient adaptive bitrate streaming protocol for \ac{5G} mmWave networks. To understand the impact of \ac{5G} network variability on the \ac{QoE} and device energy consumption, we perform a thorough pilot study (\S\ref{section:pilot}) over a simulated \ac{5G} mmWave network using the $\mathtt{ns3-mmWave}$~\cite{Mezzavilla2018} simulator framework. The current protocols for adaptive bitrate streaming attribute a higher weight to the application layer playback buffer length than the user's instantaneous received signal strength. This inevitably manifests in the increased energy consumption by accessing the network under poor channel conditions (\S\ref{section:pilot}), thereby asserting that there is room for improved energy efficiency by tuning the playback buffer length as well as the bitrates of the video downloads to network variations in the underlying \ac{5G} mmWave networks. The network throughput needs to be predicted for synchronizing the video downloads to the network changes. To address the issue related to the model training for throughput prediction, we use an innovative approach in this paper by applying the \textit{Transfer Learning}~~\cite{zhuang2020comprehensive} technique. Transfer learning is used to compensate for the limited availability of the 5G network throughput traces. For this, we first train a recurrent neural network with real-world 5G datasets~\cite{Narayanan2020, Narayanan2021, Raca2020_1} and then retrain the learned model with data collected from extensive simulations of a 5G mmWave network. The simulations help us capture every little nuance associated with the functioning of the 5G mmWave network while also allowing us to vary the number of active users in the network and observe the corresponding performance over the video streaming application. Based on the predicted future throughput, we design a reinforcement learning (RL) mechanism which dynamically tunes the maximum playback buffer length and the bitrate for the next video chunk to the network throughput variation, intending to maximize the \ac{QoE} and minimize the \ac{UE}'s energy consumption. The buffer-length and bitrate adaptation engines operate using $A3C$ \cite{mao2017neural}, a state-of-the-art actor-critic RL algorithm, and run asynchronously concerning one another. We train individual prediction modules over a large corpus of collected data traces and evaluate EnDASH-5G using an emulation environment. The source code of EnDASH-5G has been made publicly available\footnote{\url{https://github.com/arghasen10/endash} (Accessed:\today)}.

EnDASH-5G has been compared in terms of \ac{QoE} and energy consumption with state-of-the-art \ac{ABR} schemes, such as BOLA~\cite{Spiteri2016}, Fast-MPC \& Robust-MPC~\cite{Yin2015}, and  Pensieve~\cite{mao2017neural}. Our evaluation shows that in comparison to existing \ac{ABR} algorithms, EnDASH-5G significantly improves the energy savings in 5G mmWave users by 30.5\% (\S\ref{section:evaluation}). Energy saved from playing a $100$ second video using EnDASH-5G can be used to gain an additional $42$ seconds of video playback time compared to other existing \ac{ABR} algorithms.

	\section{\textbf{Background and Related Work}}\label{sec:related_work}
In this section we discuss the energy consumption in 5G \ac{NR} \acp{UE} and subsequently discuss the existing \ac{ABR} video streaming algorithms.
\subsection{UE Energy Consumption Model in 5G}
\begin{figure}[t]
	\centering
	\includegraphics[trim = {1cm 0cm 1cm 0cm}, width=0.3\textwidth]{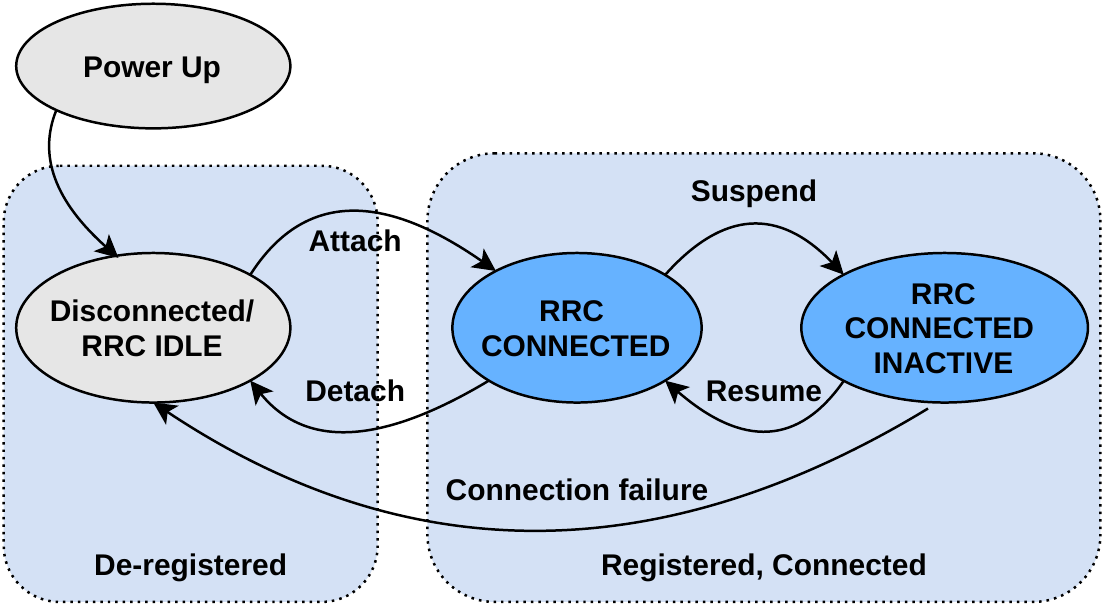}
	\caption{User-Equipment (Mobile Phone) RRC State Machine for 5G \ac{NR}\cite{RRC2017}}
	\label{fig:RRC_state_machine}
\end{figure}
\indent Energy management in 5G NR user equipments is governed by a three-state \ac{RRC} state machine as shown in Fig. \ref{fig:RRC_state_machine}. The three states are: 
\begin{enumerate*}
\item an \ac{RRC} CONNECTED state in which there is an active data transmission or reception,
\item an \ac{RRC} IDLE state with no data activity, and
\item an \ac{RRC} CONNECTED INACTIVE state between two successive active transfers.
\end{enumerate*}  
The  RRC CONNECTED INACTIVE state was introduced in 5G to reduce the control plane latency and  the energy required for transitioning from \ac{RRC} IDLE to \ac{RRC} CONNECTED state. So, in the RRC CONNECTED INACTIVE state, the \ac{UE} context is maintained both by the network and the \ac{UE}. Simultaneously, the connections between the access and core networks are also maintained. %An important point to note is that if a video is downloaded during poor connection qualities, then the smartphone spends a longer time in the \ti{CONNECTED} state leading to higher energy consumption \cite{Frenger2011}.
%\indent Most existing \ac{ABR} video streaming algorithms primarily focus on improving \ac{QoE} while paying little attention to energy savings. However, incorporating energy efficiency in video streaming will prolong the battery life of smartphones, which in turn will serve to improve the user experience.
%\indent \ac{4G} LTE smartphones are designed to maintain network connectivity using a \ac{RRC} state machine with two states: \textit{CONNECTED} and \textit{IDLE} \cite{Huang2012}. With no active transmission, the \ac{UE} is in the low power \ti{IDLE} state where no radio resource is assigned.  Once a packet arrives, the  \ac{UE} jumps to the high power \ti{CONNECTED} state, in which radio resources are assigned and data transmission takes place.  To reduce the incumbent delay and energy consumption associated with the state promotion, the \ac{UE} waits for a duration called tail time in the \ti{CONNECTED} state before returning to the \ti{IDLE} state even after packet transmission is over. To save energy in the tail period, LTE uses  \ac{DRX} during which the cellular interface periodically monitors the control channel for incoming packets and then goes to sleep ~\cite{Huang2012}.
\subsection{Related Work} 
The existing works on ABR streaming can be explored from two different directions -- works that focus on improving the video \ac{QoE} and works that target reducing the on-device energy consumption during video streaming. 
\subsubsection{\textbf{Improving QoE}}
 In~\cite{Colonnese2022}, the authors propose to improve the \ac{QoE} of video streaming through DASH-aware bandwidth allocation. ~\cite{Triki2020,Marai2018} use client-server cooperation to select video streaming bitrates and improve the \ac{QoE}. In contrast, our work assumes that no server or \ac{BS} side information is available at the client. So, the objective is to optimize the \ac{ABR} streaming at the client using only the client-side information.\\
\indent 
%\ac{ABR} video streaming algorithms choose the bitrates adaptively. 
Buffer-based \ac{ABR} algorithms like  BOLA select optimal download bitrates as a function of buffer occupancy~\cite{Huang2014, Spiteri2016}. While MPC~\cite{Yin2015}, Pensieve ~\cite{mao2017neural}, Oboe~\cite{Akhtar2018}, etc., select bitrates as a function of buffer length and network throughput~\cite{Yin2015,Jiang2014,Sengupta2018,Xu2015,Mehr2019,mao2017neural,Sani2020, Akhtar2018}. Of these, Pensieve~\cite{mao2017neural}, Oboe~\cite{Akhtar2018}, and HotDash~\cite{Sengupta2018} use deep \ac{RL} algorithm for optimal bitrate selection to maximize over a \ac{QoE} metric. Several works have used bandwidth prediction~\cite{Bentaleb2019,Raca2020,Raca2019,Raca2018_2,yue2018linkforecast} or predicted cellular network throughput  for improved bitrate selection~\cite{Raca2020,Raca2019,yue2018linkforecast,Raca2017,Raca2018_2,Raca2018_3}. The ABR streaming algorithm in \cite{Sani2020} has been trained using the combined data generated from several well-known ABR algorithms and, therefore, combines the characteristics of all these algorithms. The authors in \cite{Mehrabi2019} jointly optimize  the \ac{QoE} of mobile clients and the backhaul traffic for \ac{DASH} by integrating renewable energy into green mobile edge computing (GMEC). These works~\cite{Tuysuz2020,Yin2015,Jiang2014,Sengupta2018,Xu2015,Mehr2019,mao2017neural,Sani2020,Raca2017,Raca2018_2,Raca2018_3} improve the \ac{QoE} of users by optimizing the chunk bitrates, while paying little or no attention to improve device energy consumption.

%\indent In contrast, in \cite{Mondal2020}, authors propose to 
%% under mobility. In \cite{Mondal2020} conditions in \ac{4G} LTE networks. \\
%\indent In this work, we aim to improve video user's energy consumption over cellular networks while not compromising on \ac{QoE} by tuning playback buffer size to network throughput. Hence, the proposed algorithm should use cellular network throughput prediction.
%% \niloy{Does this mean at original level, high definition pictures cannot be seen while here high definition pictures can be seen - this somehow have to come in the study.} \cite{Siris2014} uses mobility and throughput prediction to take decisions on prefetching while moving through a \ac{WiFi}-cellular network.
% However, unlike our EnDASH algorithm, none of the works consider the unique situation of co-existence of different technologies and frequent handover from one technology to another for throughput prediction.\\ %In this work, we aim towards using the predicted cellular network throughput to improve energy consumption of \acp{UE}.\\
\subsubsection{\textbf{Reducing Energy Consumption}}\label{sec:lit_survey_energy} Several works have investigated the energy consumption of mobile phones independently of \ac{QoE} or \ac{ABR} streaming algorithms.   GreenTube in \cite{Xin2012} proposes to tune cache management to user behavior and network conditions. The energy-aware QoE-driven bitrate adaptation algorithms in \cite{Araujo2019, yue2020energy}, uses the remaining energy of the smartphone. In \cite{chen2020context} authors show the importance of the user's context in QoE estimation as well as the battery life and have designed a context-aware and energy-aware video streaming. \cite{liu2021accelerating} leverages Deep Reinforcement Learning-based approaches for finding the optimum policy to minimize the accumulated energy consumption of each \ac{BS} with minimized video playback interruptions. EVSO in \cite{Park2019} proposes a new perceptual similarity calculation method that uses the information generated by the video encoder and adjusts the frame rate according to the degree of motion intensity. This implementation is more focused on the upstream service. 
% Authors in \cite{Atawia2018} show utilization of Stochastic Modeling for energy-efficient video streaming with fewer resources while promising QoS satisfaction under network uncertainties. 
A popular method to reduce energy consumption in mobile phones is to optimize the tail energy, which is achieved in ~\cite{Yang2018} by either prefetching or delaying packets. The authors in \cite{Narayanan2021} proposed an energy-saving mechanism that opportunistically switches the network interface from 5G to 4G when the 5G mmWave network condition worsens. Such vertical handovers incur increased signalling message exchanges, as well as increased energy consumption~\cite{Mondal2020}. It can also cause ping pong vertical handovers between 4G and 5G during incessant network quality fluctuations. The consequent impact on energy consumption may be non-negligible. \\
% \cite{GunerArxiv2018} has proposed power optimization by controlling application-layer parameters such as number of parallel transfers per file, and the number of concurrent fie transfers. \\
%\paragraph{\textbf{\ac{QoS} provisioning for video traffic}} 
% Application layer protocols which prefetch and store video chunks so that the  video quality remains unaffected during poor network conditions are  proposed in \cite{Sengupta2018,Xu2015,Siris2014}. The proposed algorithm, EnDASH, predicts the cellular network throughput over a finite future time window. It chalks out a video segment download schedule and the optimal bitrates for the segments to be downloaded. 
\indent The question that remains to be answered is --  How can the \ac{QoE} of users and the corresponding energy consumption be jointly optimized? A potential answer to this question has been provided in our previous work~\cite{Mondal2020}, which proposed a method and system to improve the energy consumption of 4G smartphones during video streaming by tuning the video download sessions to the 4G network connection quality fluctuations. Tuning the bitrate to connection fluctuations reduces the CONNECTED state dwell time and, hence, the energy consumption while the bitrates are selected to maximize the \ac{QoE} of the user. Although \cite{Mondal2020} achieves energy savings, it does not investigate the implication of using the EnDASH algorithm in 5G mmWave networks, which is addressed in this work.

	\section{\textbf{Pilot Study}}\label{section:pilot}
To analyze the impact of network parameters on video streaming applications over 5G mmWave networks, it is essential to understand the complex interactions among different network-related parameters (such as received signal strength, handovers, etc.) on one hand and device/application-related parameters (such as UE speed, application throughput, device's energy consumption, video playback buffer length, video playback quality, etc.) on the other hand.  In this work, we have used extensive system-level simulation-based studies to interpret the relationship between network parameters and device/application performance counters.

\textbf{Why simulation?} Ideally, in order to understand the effect of network behavior on the performance of mobile devices,  extensive measurement-based studies on real 5G mmWave networks are desirable. However, the 5G technology remains within the development phase, and, thus, the availability of commercially deployed 5G networks is limited, particularly in the middle and low-economy countries. Although some existing works present real network 5G mmWave datasets~\cite{Raca2020_1, Narayanan2020, Narayanan2021}, these have been collected using one or two smartphones, i.e., users. As the user experienced datarate is a function of the network load, i.e., the number of users associated with a \ac{BS}, hence, it is significant to test the efficacy of any new algorithm for different network loads. However, the information associated with the network load is proprietary and is available only at the \ac{BS}.\\
\indent Using a simulation platform helps to overcome this issue wherein datasets can be generated and algorithms can be tested for any number of active users. An important advantage is that it can be executed multiple times under the same network configuration and helps regenerate the traces with different levels of detail. One can tune the network configuration and change the deployment scenario and topology. Furthermore, 1s run with a sampling frequency of $100$ (or $10$ ms time interval for generating logs) can provide 100 entries of throughput, thus producing adequate data for accurate throughput prediction. Many researchers have studied the performance of 5G in simulated platforms, such as ns-3~\cite{Mezzavilla2018, Storck2020, Shi2021}. 

%The details follow. 
% \begin{figure}[t]
%	\centering
%	\includegraphics[width = 0.5\textwidth,trim = {1cm 1cm 1cm 1cm}]{Figures/Ue_deployment.eps}
%	\caption{A Snapshot of one Simulation drop showing location of User Equipments, \ac{LTE} {eNB}, \ac{NR} \ac{gNB} cells in ns3}\vspace*{-0.6cm}
%	\label{fig:deployment_map}\vspace*{0.6cm}
%\end{figure}
%
%\begin{figure}[t]
%	\centering
%	\includegraphics[width = 0.5\textwidth,trim = {1cm 1cm 1cm 1cm}]{Figures/boX_Buffer_length_thpt_Energy_No_of_UE.eps}\vspace*{0.6cm}
%	\caption{Pilot Study: Variation of Buffer Length, Throughput, Energy/bit with different number of UE}\vspace*{0.6cm}
%	\label{fig:variation_buffer_energy}
%\end{figure}

%\subsection{Experimental Set-up}
%\subsection{Simulation Setup}

\begin{figure*}[t]%
    \begin{minipage}{0.38\linewidth}
    	\centering
    	\includegraphics[width = \linewidth]{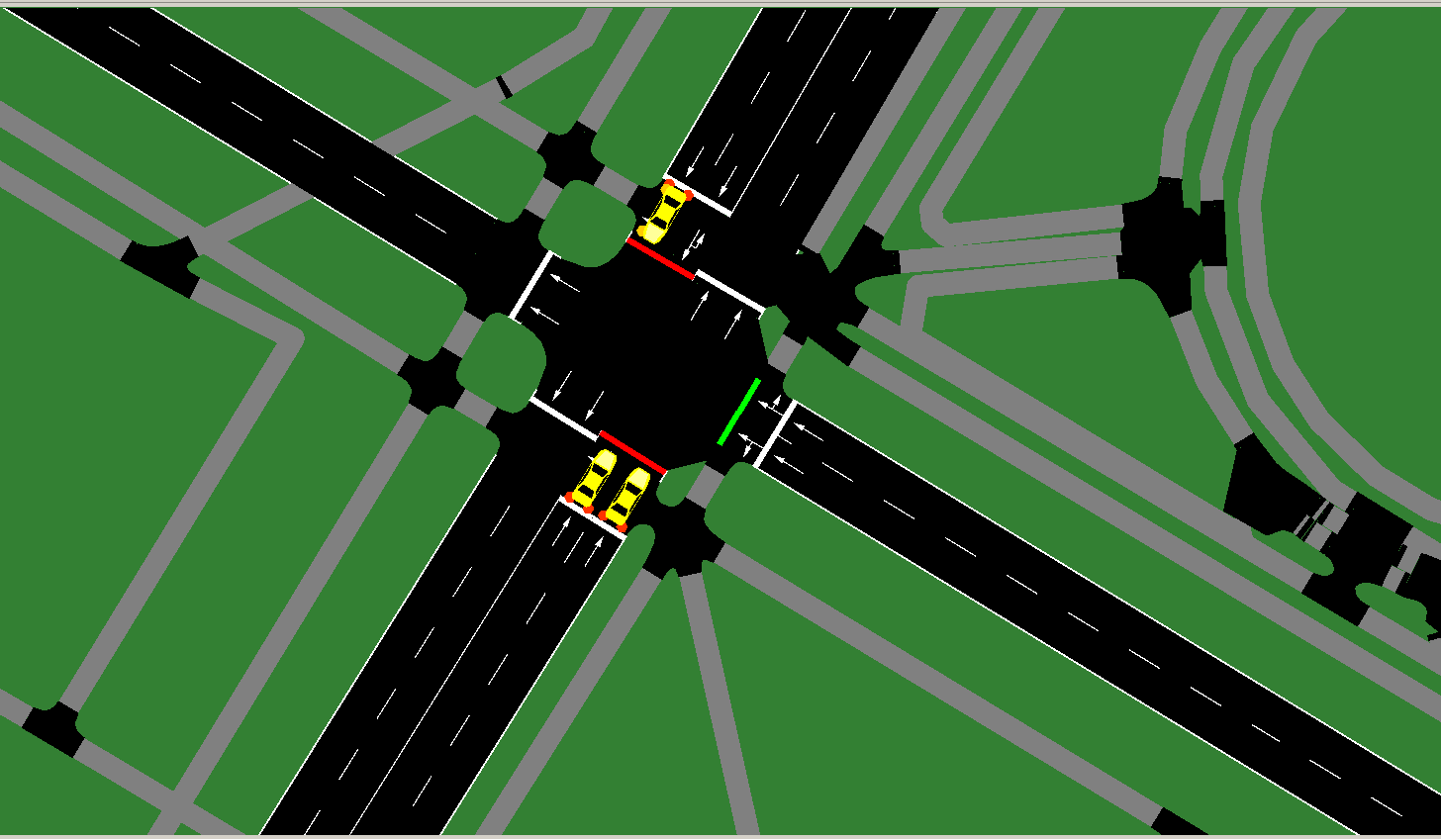}
    	\caption{Snapshot of a SUMO simulation setup of Minneapolis, USA; taken from OpenStreetMap}	
    	\label{fig:deployment_map}
    \end{minipage}
    \hspace{0.3cm}
    \begin{minipage}{0.58\linewidth}
    	\centering
    	\includegraphics[width =\linewidth,trim = {0cm  0cm 1cm 0cm}]{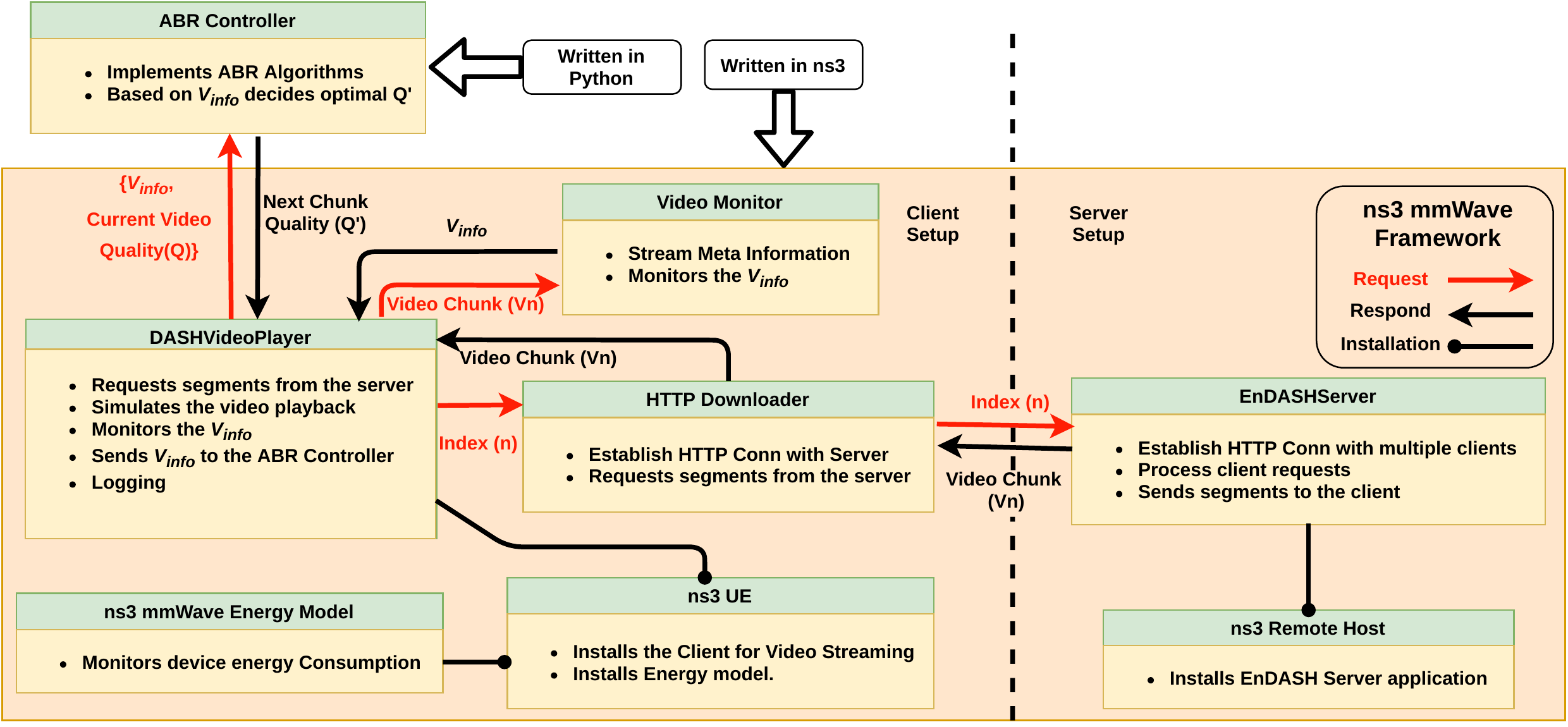}%
    	\caption{Client Server setup for video streaming simulation}	
    	\label{fig:softwareSetup}
	\end{minipage}
\end{figure*}
%\indent Besides power consumption, we have also collected extensive data on the received throughput of mobile phones in public buses, cars, and while walking across five cities of India (Kharagpur, Kolkata, Guwahati, Bengaluru, Malda). This dataset also includes data collected while travelling on highways. We have used workloads of 6Mb, 100Mb, 1GB file download, as well as of video streaming using Netflix, Hotstar, SonyLiv and Amazon Prime. This has allowed a detailed energy profiling of the smartphones. The entire corpus of collected data traces amounts to more than 50GB and has been collected over a period of eleven months.
\subsection{Experimental Setup}
\label{simu}
In this section, we explain the simulation setup for a DASH-based video client player running in a 5G \ac{UE}, which is connected to a 5G mmwave network. The setup can be broadly divided into three parts - a) a 5G mmWave network, for which we have used the $\mathtt{ns3-mmWave}$ module in \cite{Mezzavilla2018}, b) a video client player, which we have implemented in a Python setup, and c) an HTTP based DASH video streaming CDN server that hosts the video segments.
%a \ac{CDN} server. The network implementation has been done in ns3 -- which is a popular discrete event network simulator. In particular,  Certain functions concerning the video download at the \ac{UE}, such as the implementation of the algorithms has been done in a Python setup.  has also been implemented. The implementation details follow. 
%The complete experimental setup is discussed in detail in this section.
%uses baseline ABR Algorithms to predict the optimum chunk bitrate and then sends the rendered video to client side. 

\subsubsection{ns-3 Implementation of 5G Network}
%5G was designed to operate in the high-frequency mmWave range. To enable seamless connectivity with low overhead at such high frequencies, 
5G mmWave communication has two levels of access points - a) low power 5G \acp{gNB}, which operate at the higher 28 GHz frequency range, and execute the data exchange between 5G \ac{NR} devices, and b) 4G \ac{LTE} \acp{eNB}, which operate in sub-7GHz range, and oversee the communication between the 5G gNBs. Accordingly, in our simulation scenario, we have deployed a single LTE eNB at the center while $10$ gNBs are distributed uniformly within a  1km$\times$1Km square area. The detailed simulation parameters are provided in Table 1. To deploy the UEs, we have first collected a real-world map of a one square kilometer area of Minneapolis, USA, using OpenStreet Map~\footnote{https://www.openstreetmap.org}. We have then converted it into a native $\mathtt{xml}$ format using \ac{SUMO} and imported it into our  ns-3 simulation. \ac{SUMO} captures the real-world traffic pattern, which helps in analyzing the effect of vehicular mobility on 5G user throughput. The average velocity of the \acp{UE} has been varied from the pedestrian speed of $5$km/h to that of high-speed vehicles of $65$km/h.  A snapshot of the \ac{SUMO} generated simulation map is shown in Fig.~\ref{fig:deployment_map}.\\
\indent To capture the variability in network conditions, we have identified three parameters - 1) the number of \acp{UE}, 2) the \ac{UE} speed, and 3) the number of buildings deployed inside the simulation area. The number of \acp{UE} is used to vary the network load. Variation in the \ac{UE} speed models the frequency dispersion of the wireless channel. Changes in the number of buildings in a given area cause the multipath fading to vary. We have generated 27 scenarios considering different values of these parameters as shown in Table \ref{tab:simulation_setup}. We have run several simulation instances or drops of each of these scenarios to ensure a faithful averaging over a wide range of variations in network conditions and wireless channel behavior.
%\begin{figure*}[t]%
%	\centering
%	\subfigure[A Snapshot of one Simulation drop showing location of User Equipments, \ac{LTE} {eNB}, \ac{NR} \ac{gNB} cells in ns3]{%
%		\label{fig:deployment_map}%
%		\includegraphics[width = 0.48\textwidth,trim = {1cm 0cm 0cm 0cm}]{Figures/Ue_deployment.eps}}%
%	\hspace{0.1cm}
%	\subfigure[]{%
%		\label{}%
%		\includegraphics[width = 0.48\textwidth,trim = {0cm  0cm 1cm 0cm}]{Figures/boX_Buffer_length_thpt_Energy_No_of_UE.eps}}%%
%	\caption{Pilot Study and Observations}\vspace*{-0.5cm}
%\end{figure*}
\begin{table*}[ht]
	\renewcommand{\arraystretch}{1.2}
	\scriptsize
	\centering
	\caption{Simulation Parameters}
	\begin{tabular}{|p{5cm}||p{2cm}||p{5cm}||p{4cm}|}
		\hline
		\textbf{Parameter} & \textbf{Value} & \textbf{Parameter} & \textbf{Value} \\ \hline
		Bandwidth of mmWave eNBs & 1 GHz & mmWave carrier frequency & 28 GHz \\ 
		mmWave transmission power & 30 dBm & Bandwidth of the LTE eNB & 20 MHz \\ 
		LTE carrier frequency & 2.1 GHz & LTE Downlink transmission power & 30 dBm \\ 
		LTE Uplink transmission power & 25 dBm & Noise Figure & 5 dB \\ 
		eNB Uniform Planar MIMO array size & 8× 8 & UE UPA MIMO array size & 4 × 4 \\ 
		eNB scanning directions & 8  & UE scanning directions & 16  \\ 
		Sounding Reference Signal (SRS) duration & 10 μs & Period between SRSs & 200 μs \\ 
		Radio Link Control (RLC)buffer size & 10 MB & One-way delay on X2 links & 1 ms 	\\ 
		UDP payload size & 1024 bytes & UDP packet inter-arrival time & {20, 80} μs \\ 
		One-way Mobility Management Entity (MME) delay & 10 ms  & ABR Algorithms for DASH application & BOLA~\cite{Spiteri2016}, Pensieve~\cite{mao2017neural}, Robust-MPC~\cite{Yin2015}, Fast-MPC~\cite{Yin2015} \\
		\hline 
	\end{tabular}
	\label{tab:simulation_parameters}
\end{table*}
%This simulation platform is being used for generating simulation traces with video streaming applications having different ABR algorithms such as Bola, Pensieve, Robust-MPC, Fast-MPC. We have used a square platform of 1km x 1km and deployed 10 gNodeBs along with one single LTE eNodeB cell.  
	
Each UE in our simulation has an ongoing video streaming application session. The throughput data of the mobile devices have been collected for these sessions only. Now throughput varies with the network quality variations. Correspondingly, the RRC CONNECTED state dwell time of the devices and, hence, their energy consumption also changes, as shown in Fig.\ref{fig:variation_buffer_energy}. To measure the energy consumed by the 5G NR devices, we have implemented an energy module in ns-3, which has been integrated with the $\mathtt{ns3-mmWave}$ module. The implementation details of this energy module are available in \cite{Sen2021}. During an ongoing video session, the data recorded at each UE includes - (a) network-related parameters, such as the network throughput, \ac{RSSI}, \ac{RSRQ}. \ac{RSRP}, \ac{MCS}, number of handovers, and data state (CONNECTED or IDLE), and (b) UE parameters, such as distance from the \acp{gNB}, device speed, energy consumption per bit.
%\indent For this, we have modelled a Python DASH video streaming server that emulates the \ac{CDN} server. The DASH client application is installed over the $\mathtt{ns3-mmWave}$ \ac{UE}. It fetches the video segments from the Python server using one of the following \ac{ABR} streaming algorithms - (i) Pensieve~\cite{mao2017neural}, (ii) Fast-MPC~\cite{Yin2015}, (iii) Robust-MPC~\cite{Yin2015}, (iv) BOLA~\cite{Spiteri2016}, (v) the proposed EnDASH-5G. 
\begin{table}[t]
	\scriptsize
	\centering
	\caption{Simulation Scenarios}\label{tab:sim_scenarios}
	\begin{tabular}{|p{2cm}||p{2cm}|p{2cm}|}
		\hline
		\textbf{Number of UE} & \textbf{Speed (km/h)} & \textbf{Buildings Deployed}\\
		\hline \hline 
		1 & 5 &  0\\ \hline
		10 & 30 & 10\\ \hline 
		20 & 65 & 20\\ \hline 
		
	\end{tabular}\vspace*{-0.5cm}
	\label{tab:simulation_setup}
\end{table}

\subsubsection{Video Streaming Setup} \label{ssxn:serverClient}
To interpret how throughput fluctuation affects the video streaming in 5G mmWave communication, we have implemented a \ac{DASH}  \textit{server-client} system operating in the downlink. 
The video streaming setup is shown in Fig. \ref{fig:softwareSetup}. %It has two main functional blocks: a client and a server. 
The client module, which executes the majority of the functions in the setup, has been designed partially in both $\mathtt{ns3-mmWave}$ and using a Python setup.  The server, on the other hand, developed in $\mathtt{ns3-mmWave}$, implements an HTTP server to emulate the functions of a CDN server.
%contains most of the functionality of the setup. It collects and maintains the information required by the adaptation algorithm, such as the buffer level, throughput information. It uses baseline ABR algorithms to decide the optimal video bitrate for video chunk download and send periodic requests to the server module. Instead of real video traffic, simulation traffic is used to capture the streaming behavior. 
The details of the client-server setup follow.\\

\noindent\textbf{Video Streaming Client:} \label{sssxn:client}
The $\mathtt{DashVideoPlayer}$ in Fig. \ref{fig:softwareSetup} is the  $\mathtt{ns3-mmWave}$-based  application which executes the client functionalities as explained next.
\begin{enumerate}
\item The client first initiates a streaming session by establishing an HTTP connection with the video server, using the \textbf{HTTP downloader} sub-module, which has been implemented in $\mathtt{ns3-mmWave}$.
   % \item The client first initiates a streaming session by establishing an HTTP connection with the video server, using the \textbf{HTTP downloader} sub-module. This sub-module is implemented in $\mathtt{ns3-mmWave}$. The client inherits the {ns3} $\mathtt{HttpClientBasic}$ application to communicate with the HTTP Streaming Server. 
    \item The client simulates the video playback by monitoring the playback buffer status and the throughput. This is done by the \textbf{Video Monitor} sub-module, which monitors and stores video session-related information, such as playback buffer occupancy, playback buffer length, rebuffer time, last chunk quality, and average throughput, etc. in the metadata object $V_{info}$.
    \item The DASH video player sends the $V_{info}$ to the \textbf{ABR controller}. The controller runs an ABR algorithm, such as Pensieve or BOLA, which determines the optimum quality for the next video chunk and sends it back to the DASH video player. The ABR Controller module is not implemented in ns-3. It is a separate Python server hosted on the same local machine where the ns-3 simulation runs.
    \item The client maps the next chunk download quality to the corresponding bitrate and requests the streaming server for the video segment through the \textbf{HTTP Downloader} module. It stops sending requests to the server when the streaming session is over
\end{enumerate}
\noindent{\textbf{Video Streaming Server:}}
Videos are "Dashified" and stored in a streaming server. The EnDASH Server manages connection with  clients and processes client requests. Upon receiving the client's request, the server responds with the requested video chunk in the requested quality.  We first explain the video dashification and then the  EnDASH server implementation.
\begin{enumerate}
    \item \textbf{Dashifying videos:} In this work, we have used $56$ DASHified videos~\cite{Mondal2020} and streamed them over the simulation setup. Dashifying implies the conversion of the video to a form suitable for playing it using a DASH player. Essentially, the videos have been split into segments with fixed duration, and then each segment is stored at multiple qualities. The length of each video chunk is kept as $8$sec, while the chunk size, in terms of bytes, depends on the encoding bitrate value. Each segment has been stored in the CDN server in the following encoding - $\left[144p,270p,360p,480p,570p,720p\right]$.  We have used the open-source `ffmpeg' non-interactive video manipulation tool to dashify the videos. All the videos are stored at the server and have been streamed over  the  different simulation scenarios outlined in Table~\ref{tab:simulation_setup}.
    \item \textbf{Implementing the server:} To implement the server, we have developed $\mathtt{EnDASHServer}$ -- an ns-3 based application that is installed over a remote host.
    In each incoming request, the server responds with the requested segment based on the information it receives about the segment length or the number of bytes it has to send. 
\end{enumerate}
\subsection{Results and Observations}

\begin{figure}[t]%
	\centering
	\includegraphics[width = \linewidth,trim = {0cm  0cm 1cm 0cm}]{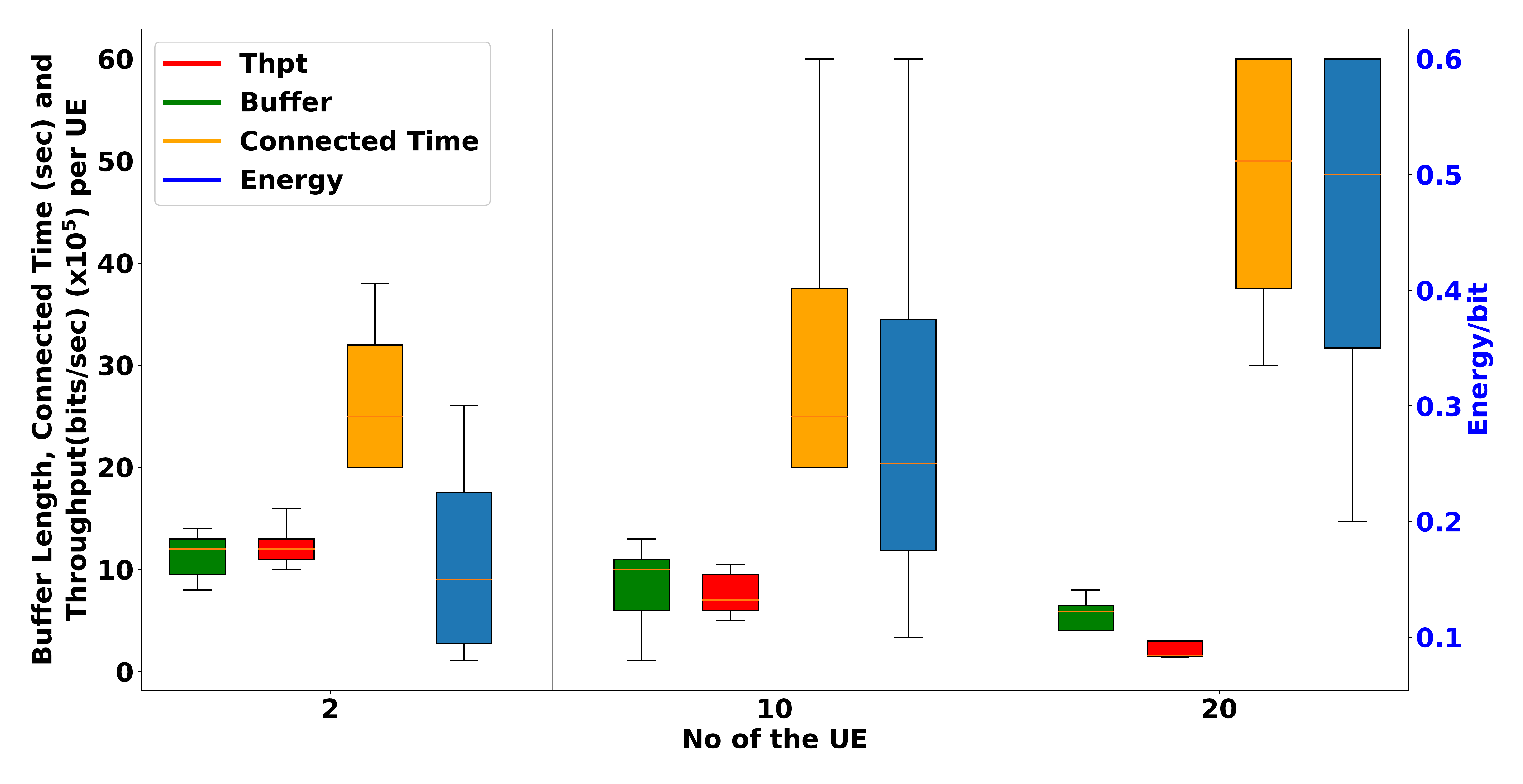}%
	\caption{Variation of Buffer Length, Connected state dwell time, Throughput, Energy/bit, with the number of UE; ABR streaming algorithm used - BOLA.  }	\label{fig:variation_buffer_energy}
\end{figure}
\begin{figure*}[!t]
	\centering
	\subfigure[]{%
	    \centering
	    \label{fig:energy_rssi}%
		\includegraphics[width=0.24\textwidth,keepaspectratio]{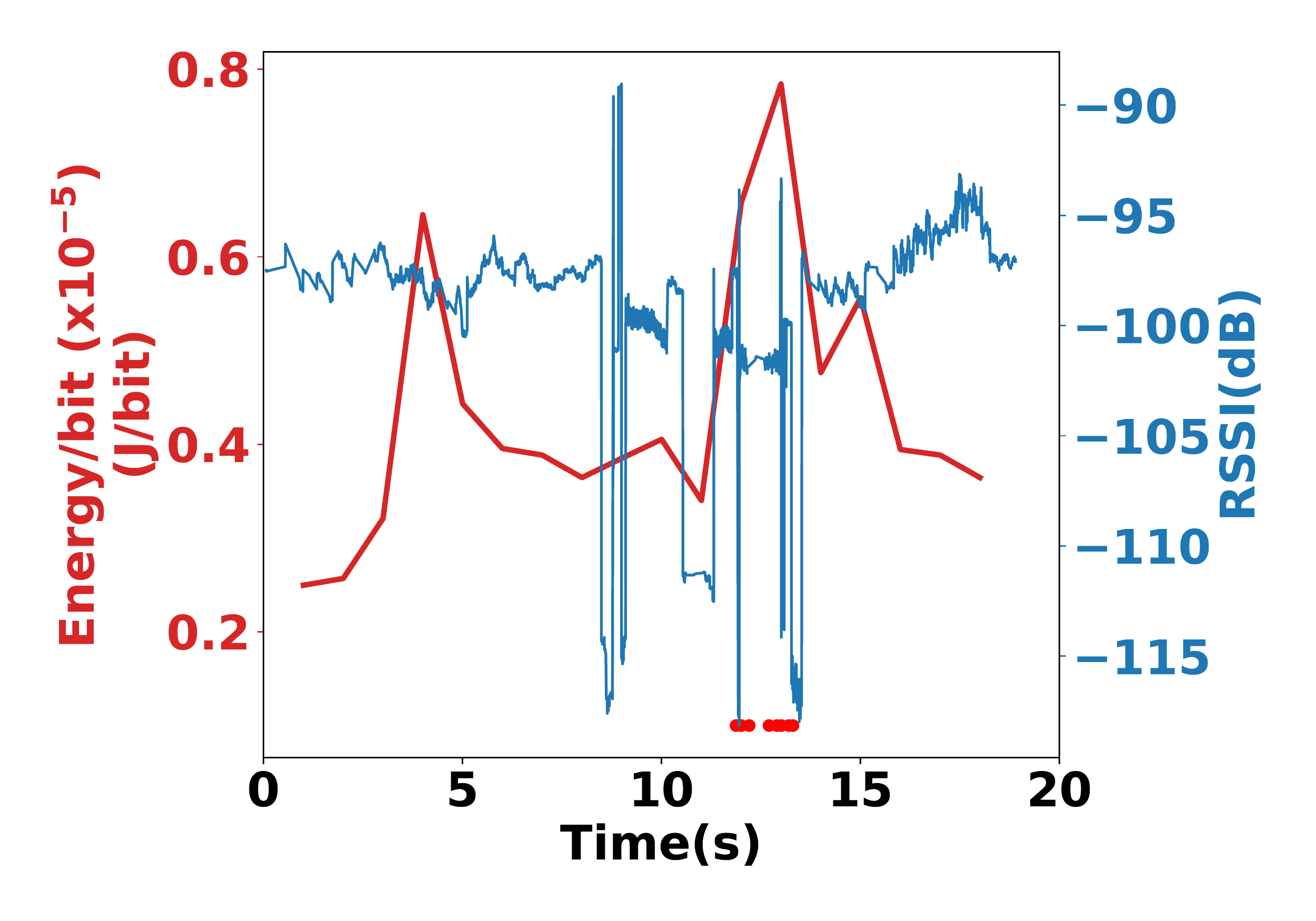}}\hfil
	\subfigure[]{%
	    \centering
		\label{fig:throughput_rssi}%
		\includegraphics[width=0.24\textwidth,keepaspectratio]{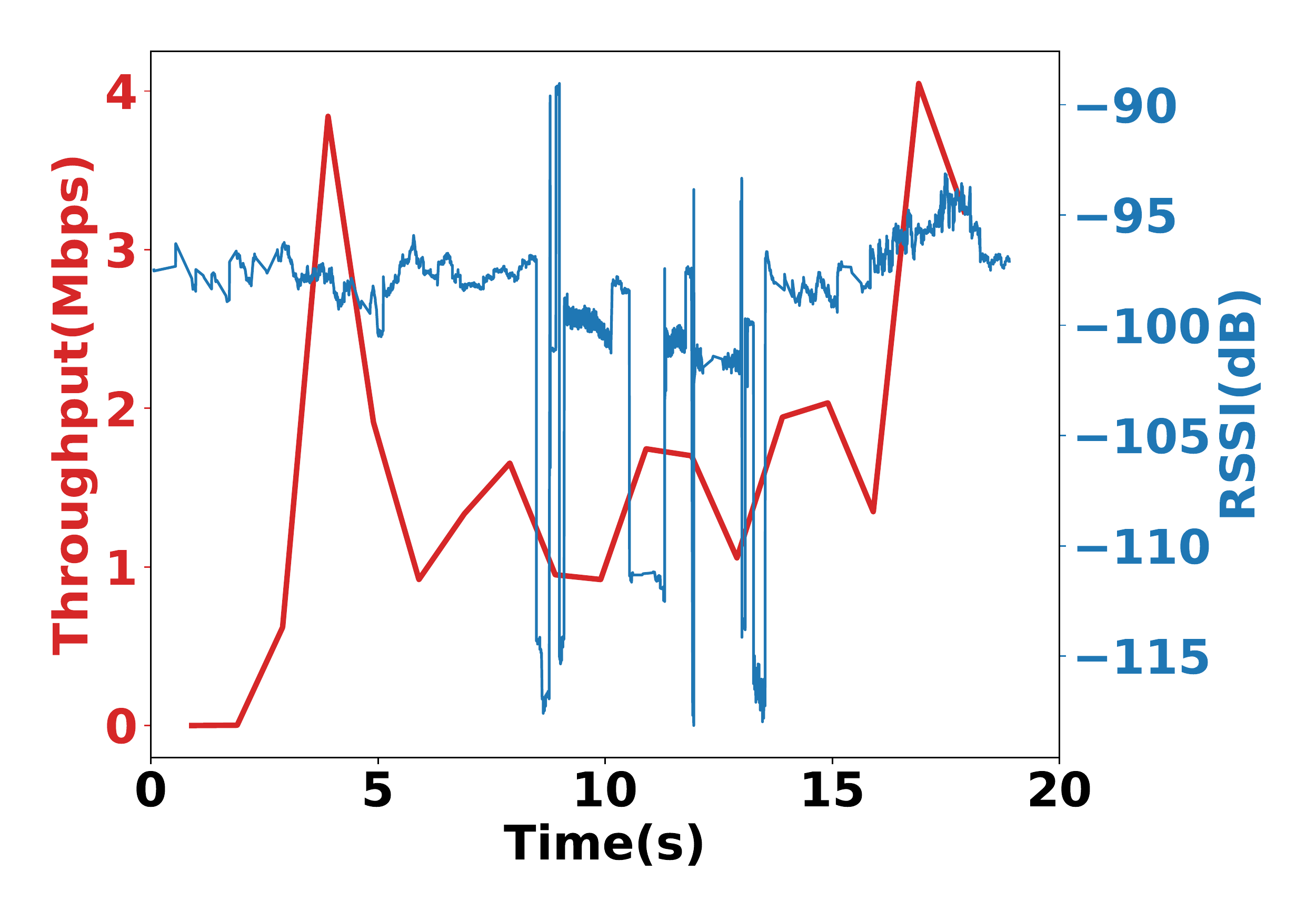}}\hfil
	\subfigure[]{%
	    \centering
		\label{fig:bitrate_rssi}%
		\includegraphics[width=0.24\textwidth,keepaspectratio]{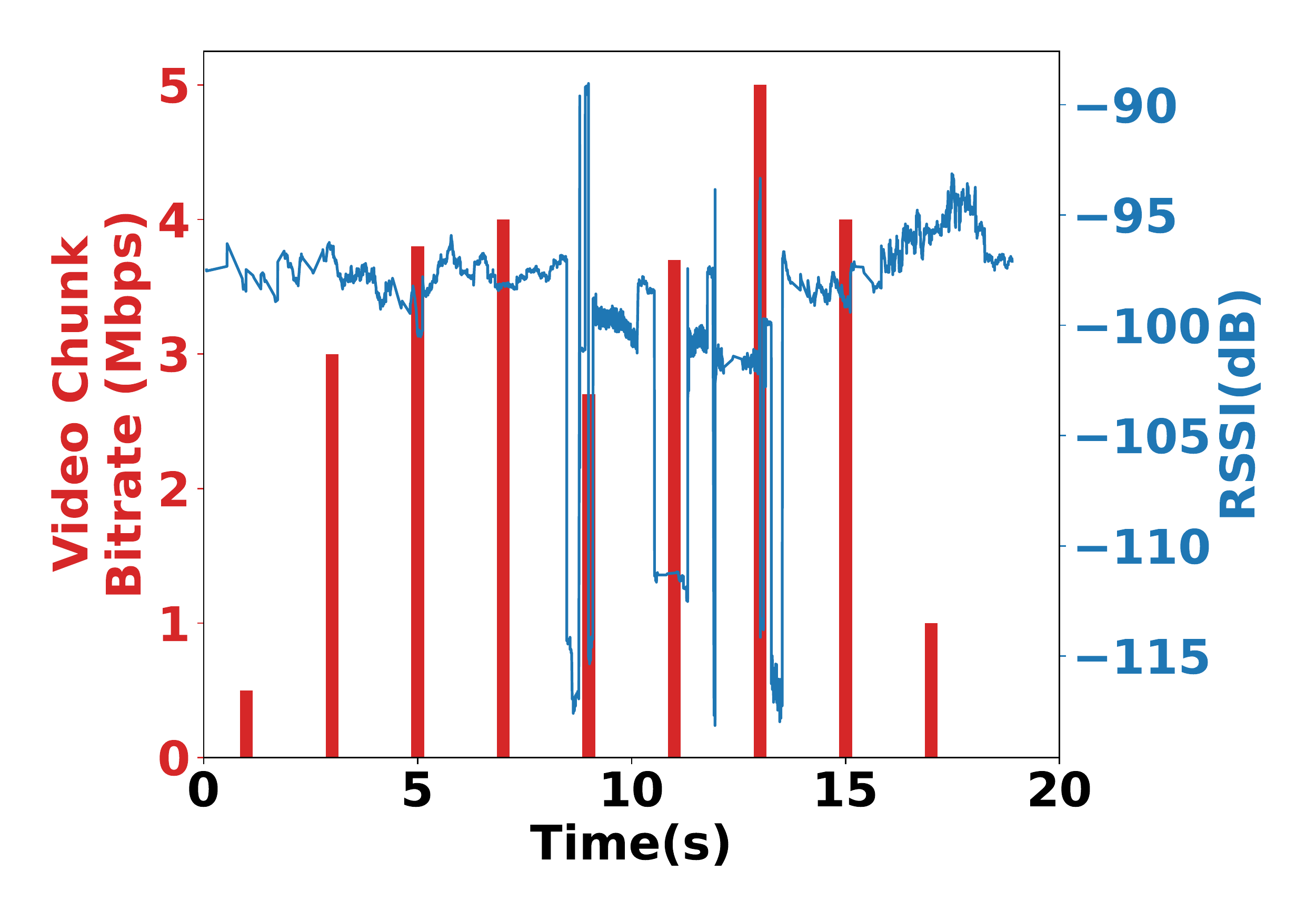}}\hfil
	\subfigure[]{%
	    \centering
		\label{fig:buffer_rssi}%
		\includegraphics[width=0.24\textwidth,keepaspectratio]{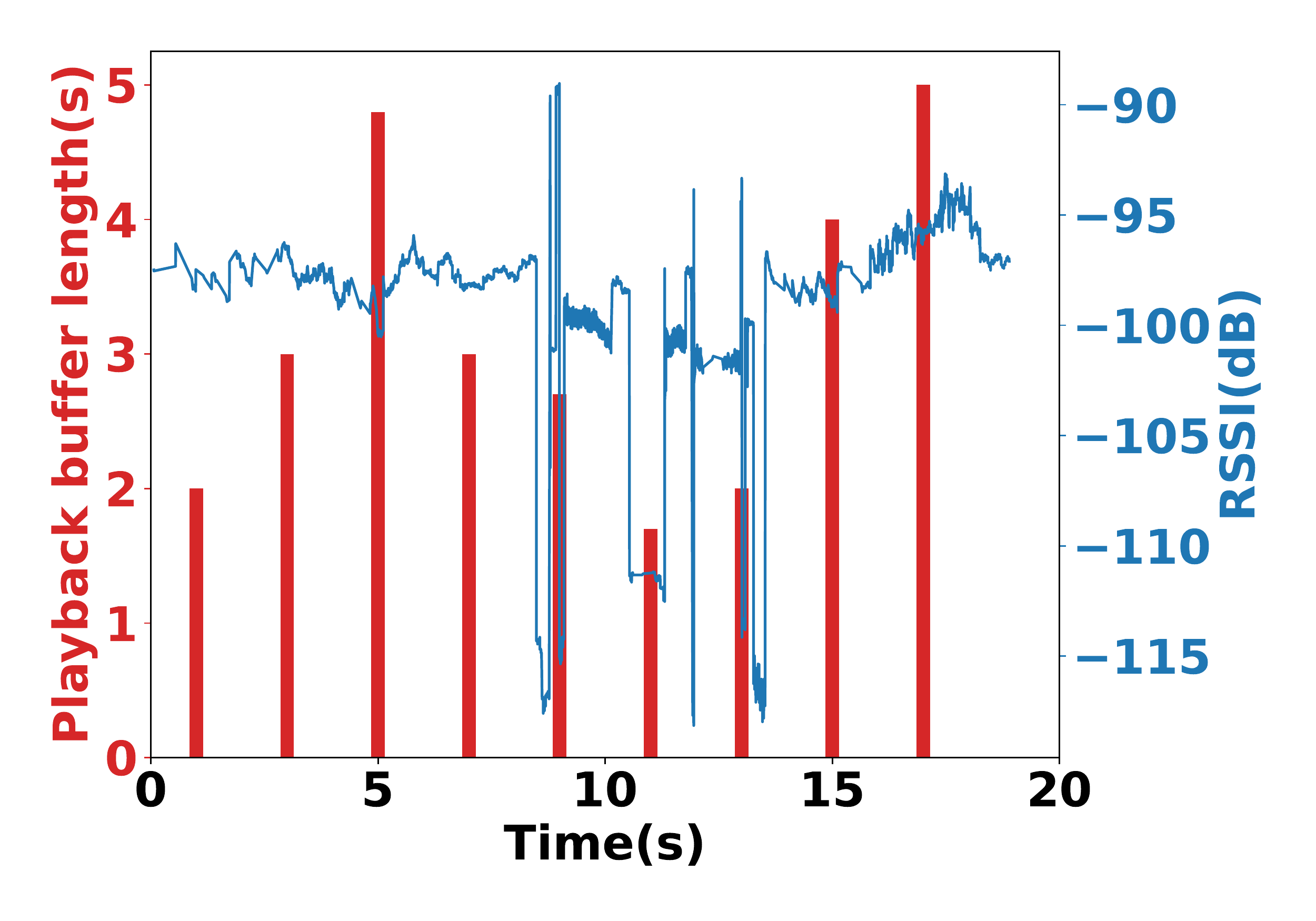}}\hfil
	\caption{Pilot Study: Fig.(a) Variation of Energy per bit with time. Fig.(b) Variation of Throughput with time Fig.(c) Variation of Bitrate with time. Fig.(d) Variation of Playback buffer length with time. The variations of all the metrics are shown against the temporal variations of RSSI.}
	\label{fig:pilot_study_results}
\end{figure*}
Using the setup in \S\ref{simu}, we have made a detailed analysis of the effect of 5G network quality fluctuations on video streaming using the log generated by the $\mathtt{DashVideoPlayer}$.\\% The energy consumption study of the UEs has been recorded using the $\mathtt{ns3-mmWave}$ Energy module~\cite{Sen2021} which is external to the video playing application.\\}
\indent Fig. \ref{fig:variation_buffer_energy} shows the variation in the network throughput, the available playback buffer length, time spent in the connected state, and the average energy consumption per bit of an individual user as a function of the total number of \acp{UE} in the network. The average speed of each \ac{UE} is 30km/h, and the number of buildings deployed in the one square kilometer simulation area is $10$. We have simulated three scenarios with two, ten, and twenty users, respectively, keeping the speed, number of buildings, and network parameters unchanged. We have run five simulation drops of each of these scenarios, and each drop has a duration of $100$ seconds. Averaging over the simulation drops yields Fig. \ref{fig:variation_buffer_energy}. Each whisker plot shows the maximum and the minimum on the two extreme sides. The upper and lower boundaries of the box represent the 75$\textbf{}^\text{th}$ and 25$\textbf{}^\text{th}$ percentile points. The line inside the box represents the median, i,e, the 50$\textbf{}^\text{th}$ percentile point.\footnote{All whisker plots in the paper follow this same convention here.} It is observed that the network throughput and the buffer length decrease with the increase in the number of \acp{UE}. At the same time, there is an increase in the energy consumption of the individual \acp{UE}.\\ %This indicates that an increase in the number of active users and, hence, the traffic load delivers a lower throughput per user, resulting in higher energy consumption. This is due to the longer time spent in the high power connected state as also reflected in Fig. \ref{fig:variation_buffer_energy}.\\
\indent An in-depth study of the plots reveals that the per-user throughput is high with fewer users in the network, which allows the application layer to download a higher number of video segments at a higher rate. This, in turn, increases the playback buffer length. \textbf{A higher rate of download implies a lower CONNECTED state dwell time}, which manifests in the lower energy consumption as seen from  Fig.~\ref{fig:variation_buffer_energy}. It may be inferred that an increase in the number of users increases the network load thereby dividing the network resources among more users. This causes lower throughput, shorter playback buffer length, and higher energy consumption per user.

\indent Of all the simulation drops, we have shown in Fig. \ref{fig:pilot_study_results} the snapshot of the worst-case scenario with 20 \acp{UE}, each moving with an average speed of 65km/h, in a scenario with 20 buildings deployed. This snapshot has been generated with BOLA~\cite{Spiteri2016} as the ABR streaming algorithm. It shows the variation of energy consumption per bit, video segment download throughput, video chunk bitrate, playback buffer length, received signal strength (in blue), and the handover events (shown in red dots) of a typical \ac{UE}. In line with \cite{Mondal2020}, the trace in Fig.~\ref{fig:pilot_study_results} shows that the volume of video segments downloaded is not always commensurate with the connection quality, which in this work has been quantified by the \ac{RSSI}. As can be seen in Fig.~\ref{fig:pilot_study_results}(a), in the event where there are frequent handovers between 11 seconds and 14 seconds, the \ac{RSSI} shows high variability, i.e., the channel quality worsens. This is also observed in Fig.~\ref{fig:pilot_study_results}(b). Nevertheless,  Fig.~\ref{fig:pilot_study_results}(c) shows that even at the time of handover, the \ac{UE} commits to packet download at high bit rates. This attempt to download video segments at a higher bitrate even in fluctuating throughput causes the device to download the chunk for a longer time. Hence, it stays in the high-power RRC CONNECTED state for longer. Consequently, the energy consumption increases as is reflected in Fig.~\ref{fig:pilot_study_results}(a). On the other hand, even when the channel quality is good between 4 seconds and 8 seconds, the player downloads the video segments at a lower rate, as observed in Fig.~\ref{fig:pilot_study_results}(c).\\
\indent We may, therefore, infer from Fig.~\ref{fig:pilot_study_results} that some parameters other than the channel quality also drive the QoE during video streaming. The rate at which the video buffer is filled depends on: a) the available bandwidth and b) the quality of the video segments downloaded. Once the buffer length exceeds a threshold, the download stops and restarts only when the buffer length goes below the threshold. So, if the buffer is full when signal quality improves, then the UE does not download any video packet. We can, therefore, summarize our takeaway from this pilot study as below.

\noindent $\mathrm{\mathbf{\underline{Takeaway:}}}:$ \textit{The current protocol of video download always attributes a higher weight to the playback-buffer length than the user's instantaneous received signal strength or throughput, ensuing a significantly high energy consumption.}

\indent This inference from the pilot study provides the design criteria for developing an energy-efficient video streaming mechanism for 5G systems, which is discussed next.
	\section{\textbf{EnDASH-5G System}}\label{section:sys_overview}
EnDASH-5G acts as a wrapper over \ac{DASH} enabling energy-efficient video streaming over HTTP in 5G mmWave networks. Based on the observations from the pilot study, EnDASH-5G  engages in the opportune fetching of video chunks during good channel conditions, which is facilitated by the prediction of the channel quality over a finite future time window. This reduces the CONNECTED state dwell time and eventually economizes the energy expenditure. To accommodate the higher volume of download, the core idea of EnDASH-5G is to dynamically tune the playback buffer length to the current network and wireless channel conditions, unlike existing players that use a static buffer. Once the playback buffer length is adjusted, EnDASH-5G, an ABR video streaming algorithm, decides the video chunk bitrates as a function of this decided playback buffer length and the current channel condition. %We next describe the flow of EnDASH-5G.

\begin{figure}[!t]
	\centering
	\includegraphics[width = \linewidth,trim = {1cm 1cm 1cm 1cm}]{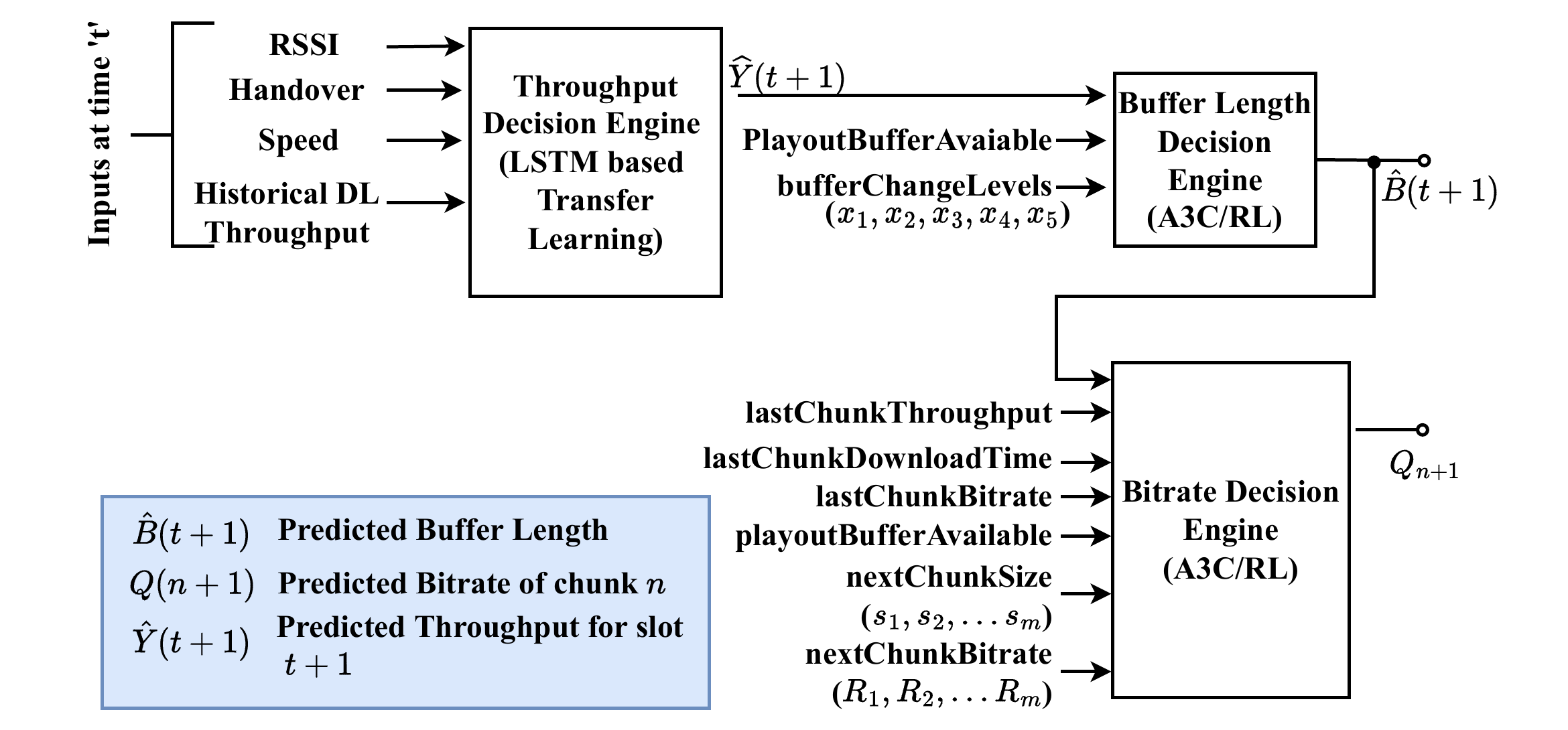}
	\caption{Composite Representation of the EnDASH-5G model; a cascaded model where the predicted throughput acts as an input network state to the Actor Critic \ac{RL} based decision engine}
	\label{fig:EnDASH-5G system}
\end{figure}

EnDASH-5G operates in a time-slotted manner, i.e., it divides the timeline into time-slots of length \mq{\mathcal{W}}~\cite{Mondal2020}.  Its dynamics move through a cascade of three engines: (i) throughput decision, (ii) buffer length decision,  and (iii) bitrate decision, as shown in Fig. \ref{fig:EnDASH-5G system}.
%To achieve energy efficiency, it executes the following: 
\begin{enumerate}
    \item At the beginning of any time slot `$t$', EnDASH-5G executes the following:
    \begin{enumerate}
        \item In the \textit{throughput decision engine}, it first \textit{predicts the average \ac{UE} throughput} in the current time slot, i.e., for the next `$t+\mathcal{W}$' seconds~\cite{Mondal2020} by exploiting the  network parameter and the downlink throughput information of the previous `$t-H$' seconds.
    \item Next, in the \textit{buffer length decision engine}, it uses this predicted throughput to \textit{estimate the optimal average playback-buffer length} for the duration of the current slot such that the energy consumption is minimized, using the an \ac{A3C} deep \ac{RL} algorithm.
    \end{enumerate}
    %There  limited availability of real-network 5G throughput traces, in this work the EnDASH-5G throughput prediction engine is trained by transferring knowledge gained during 4G throughput prediction, using a \textit{transfer learning}-based approach~\cite{zhuang2020comprehensive}. 
    
    \item Finally, within a time slot, just before downloading each video chunk, EnDASH-5G enters the \textit{bitrate decision engine} where it uses the predicted optimal buffer length as an input to another A3C deep \ac{RL} engine to predict the optimal download bitrate of the video chunk.
\end{enumerate}
The detailed functioning of these decision engines follow.

\subsection{The Throughput Prediction Engine}\label{section:thpre}
% \extension{\begin{itemize}
% 	\item	Throughput Prediction to include Transfer Learning using LSTM
% 	\end{itemize}}
 
The first step towards the implementation of EnDASH-5G is the design of an efficient throughput prediction engine for 5G cellular networks. As discussed in \cite{Raca2019,Mondal2020}, the cellular network throughput varies with a wide range of factors like -- radio signal strength at the user devices, the technology of the associated and neighboring \acp{BS}, \ac{UE} speed, the handovers between \acp{BS} of same or different technologies, the number of users being served by a \ac{BS} as well as the radio resource allocation algorithms. It is, thus, evident that there exists a non-trivial relation between the cellular network throughput perceived by a running application and the corresponding network parameters. So, traditional averaging mechanisms such as arithmetic, exponential, or harmonic moving averages used by most \ac{ABR} streaming algorithms do not perform well for predicting cellular network throughput in practice~\cite{Raca2019}. 
\subsubsection{Existing Works on Throughput Prediction and Their Limitations in the Context of 5G Networks}%\textcolor{red}{Ei subsubsection heading ta agey pathate hobe}
Several existing works in the literature~\cite{Raca2017,Mondal2020,Raca2019,Raca2020,yue2018linkforecast,Samba2017} have explored machine learning algorithms for predicting network throughput perceived by a running application. Works like~\cite{Raca2017,yue2018linkforecast,Raca2019,Mondal2020} train an \ac{RF}-based supervised learning algorithm for predicting 4G network throughput using real-network throughput traces. Deep Learning models~\cite{Raca2020} perform significantly better than the \ac{RF}-based models; however, these models need a huge training dataset to train the model properly. \\
%These trained 4G models  need to be retrained with 5G data in order to account for the additional network and context parameters which affect the 5G throughput. \argha{\textcolor{red}{ However, as mentioned earlier, the commercial deployment of 5G is still in a nascent stage. So, in this work, we train the throughput prediction engine using simulated (synthetic)  \\
\indent  As mentioned earlier, the commercial deployment of 5G is still in a nascent stage. Besides, most existing 5G throughput traces do not contain information associated with the variability in the network load, i.e., the number of active users. Hence, in this work, we have used synthetic 5G network throughput traces generated as in \S\ref{simu}. Nevertheless, there is a significant departure in the statistical characterization of simulated traces from real network traces. To accommodate for this, i.e., to imbibe the salient features of real-network throughput prediction in our model, we propose to use \textit{transfer learning}~\cite{zhuang2020comprehensive}. In transfer learning, the knowledge gained from previously learned tasks is applied to a target domain of related tasks. Interestingly, the publicly available 5G datasets~\cite{Narayanan2020, Narayanan2021, Raca2020_1} provide enough throughput traces that can be used as a source domain for the transfer learning and can be applied to a target 5G domain with a simulated dataset. To the best of our knowledge, this is the first work that uses transfer learning to predict  5G throughput using simulated traces.
 \subsubsection{Transfer Learning}
%The building block of our transfer learning approach is a deep recurrent neural network in which the first two layers are LSTM layers while the third one is a fully connected layer. 
%The basic building block of transfer learning is a deep neural network.
In transfer learning, given a source domain $D_S$ with a significant amount of data and a target domain $D_T$ with a limited amount of data, a deep neural network model $M_S$ is first trained for learning a task $\zeta_S$ in $D_S$. Subsequently, the model $M_S$ is retrained with the dataset of $D_T$ for learning a task $\zeta_T$ in $D_T$. The knowledge, i.e., the features, weights, etc., gained while training $M_S$ in $D_S$ enhances the learning of the target predictive function $\mathcal{F}_T(.)$. Here, $D_S \neq D_T$, and $\zeta_S$ may or may not be the same as $ \zeta_T$. Transfer learning effectuates a shorter training time while improving the accuracy even with a smaller dataset size in the target domain. In our work, the source domain $D_S$ corresponds to real 5G network throughput data while the target domain $D_T$ corresponds to the simulated 5G network. The source and target tasks $\zeta_S$ and $\zeta_T$ both correspond to the prediction of throughput.  So, we have $D_S \neq D_T$ but $\zeta_S = \zeta_T$.
% In transfer learning, knowledge (features, weights, etc.) from previously trained models are leveraged for introducing newer models. 
% If we have significantly more data for task $T_S$, we may utilize its learning and generalize this knowledge (features, weights) for task $T_T$ (which has significantly less data).
%\subsubsection{Source and Target Domain for the Proposed Throughput Prediction Model}
% \textbf{5G Throughput Prediction Approach}
 %generated using the setup described in  We need to select features that are common between $D_S$ and $D_T$, as fine tuning on the $D_T$ will expect the same features as in the $D_S$. We found RSSI, distfromcell, speed, throughput as common features.
%  \begin{figure}[t]
%  	\centering
%  	\includegraphics[width = \linewidth]{Figures/tf_sch2.pdf}
%  	\caption{Transfer Learning Model for the 5G Throughput Prediction}
%  	\label{fig:TL}
%  \end{figure}
 \subsubsection{Model Description and Transfer Learning}\label{sec:TL_LSTM}
% In this section, we explain the training and implementation of the transfer learning-based throughput prediction algorithm. 
The proposed throughput prediction engine, outlined in Algorithm \ref{algo_Thpt_prediction},  is designed to predict the network throughput $\mathcal{W}$ seconds into the future based on the information available in the network and user-related parameters as well as that of the downlink throughput of the previous $H$ seconds.   At the heart of this prediction engine is a  recurrent neural network with  \textbf{two layers} (128 units each) of \ac{LSTM} cells followed by a \textbf{fully connected layer} of 1 node. We have also added a dropout of 0.2 after every LSTM layer as a form of regularization. LSTM is often chosen as the preferred model for data training when the corresponding dataset is quite large. In this work, we have significantly large real 5G network as well as  simulated 5G network throughput traces available. Moreover, throughput data is in time-series format \cite{Raca2020}. Hence, we have used LSTM. The details  follow.
\paragraph{\textbf{The Source Domain}}
The first step  is training the LSTM model $M_S$ using the source domain dataset in order to learn the weights $\theta_S$. This dataset is a collection of data points $\{\mathbf{X}_S^{i} , Y_S^{i}\}$, $\forall i\in \tau_\mathcal{S}$, where $\tau_\mathcal{S}$ is the observation time. The set $\mathbf{X}_S^i = \{X^{i}_{(S,1)} , X^{i}_{(S,2)}, ... , X^{i}_{(S,n)}\}$ is the set of `$n$' Radio Channel and Location Metrics while $Y_S^{i}$ is the downlink throughput, at the time step $i$. %We first train the LSTM network using the source domain 4G-India dataset to learn the weights $\theta_S$. Subsequently, we retrain $M_S$ using the target domain dataset $\mathcal{D}_T$. These steps are explained next.
 %retunes $\theta_S$ to $\theta_T = \theta_S+\Delta\theta_S$ thereby adapting the model to the target domain. This is shown in Fig. \ref{fig:TL} % We use \textbf{LSTM} based Deep Learning model since LSTM's are good at modeling sequential/temporal data where the variable to be predicted has dependencies on historic values as well.
% The model consists of 
% Fig. \ref{fig:Transfer Learning Approach} provides the overview of the approach.
 %\subsubsection{Algorithm}
 %\textcolor{red}{This is the only subsubsection in subsection 4.2. Should it stay?}
 \begin{algorithm}[t]
 	\caption{Throughput Prediction Algorithm}\label{algo_Thpt_prediction}
 	\begin{algorithmic}[1]
 	 \Procedure{Train\_LSTM}{$\mathcal{D}_{S}$}
  	    \For{$i \in \tau_S $}
 	        \State $\{\mathbf{\psi}^i_{X,S}, \psi_{Y,S}^i \}=$ create\_samples$(\mathcal{D}_S,H,\mathcal{W})$. \label{algo_creatsourcesample}
 	        \State Find $\theta_S=\argmin{\theta_S} \mathrm{loss}(\hat{Y_S}^{(i+\mathcal{W})}, \overline{Y_S}^{(i+\mathcal{W})})$.\label{algo_learn_LSTM}
 	        \item[]
 	        \Return $\theta_S$
 	    \EndFor
 	\EndProcedure
 	\State $\mathcal{D} = \mathrm{preprocess}(\mathcal{D}_T)$.\label{algo_preprocess}
 	\State $\mathcal{D}_{norm}$ = $\frac{\mathcal{D} - min(\mathcal{D})}{max(\mathcal{D}) - min(\mathcal{D})}$.\label{algo_normalize}
 	\Procedure{Fine\_Tune\_LSTM}{$M_S, \theta_S, \mathcal{D}_{norm}$}
        \For{$i\in\tau_T$}
        \State $\{\mathbf{\psi}^i_{X,T}, \psi_{Y,T}^i \}=$ create\_samples$(\mathcal{D}_{norm},H,\mathcal{W}).$\label{algo_creattargetsample}
            \State Find $\Delta\theta= \argmin{\Delta\theta} \mathrm{loss}(\hat{Y_T}^{(i+\mathcal{W})}, \overline{Y_T}^{(i+\mathcal{W})}).$\label{algo_learn_LSTM_target}
            \item []
            \Return $\Delta\theta$.
        \EndFor
 	\EndProcedure
 	\State Generate final weights $\theta_T=\theta_\mathcal{S}+\Delta\theta$.
 	\end{algorithmic}
 \end{algorithm}
%The throughput prediction algorithm, outlined in Algorithm \ref{algo_thpt_prediction}, predicts the `\textit{average throughput}' over a finite future time window of $\mathcal{W}$ seconds based on the historic throughput and network data for the past $H$ time steps. 
To enable the training and testing using the LSTM cells,  the collected data in time series format is mapped to two data matrices $({\psi_{X,S}^i} , \psi_{Y,S}^i), \forall i$ (Line \ref{algo_creatsourcesample}). While $\psi_{X,S}^i$ represents the matrix of network parameter and location related features, $\psi_{Y,S}^i$ represents the vector of downlink throughput from $i-H$ seconds to $i-1$ seconds,  i.e., 
\begin{equation}\label{eq:createsample}
    {\psi_{X,S}^i} = \begin{pmatrix}
 			X_{S,1}^{(i-H)} & X_{S,2}^{(i-H)} & ... & X_{S,n}^{(i-H)}\\
 			X_{S,1}^{(i-H-1)} & X_{S,2}^{(i-H-1)} & ... & X_{S,n}^{(i-H-1)}\\
 			\vdots & \vdots & \vdots & \vdots\\
 			X_{S,1}^{(i-1)} & X_{S,2}^{(i-1)} & ... & X_{S,n}^{(i-1)} \end{pmatrix}.
\end{equation}
 
\begin{equation}\label{eq:createsample_thpt}
    \psi_{Y,S}^i = [Y_S^{i-H} \ Y_S^{i-H-1} \ \cdots Y_S^{i-1}].
\end{equation}
%$\psi_y^i$ represents the actual average throughput $\overline{Y}^{(i+T)}$ from timestep $i$ to timestep $i+\mathcal{W}$.
% \begin{equation}
%     \psi_y^i = \overline{Y}^{(i+\mathcal{W})} = \frac{1}{\mathcal{W}}\sum_{t = i}^{i+\mathcal{W}}Y_S^{(t)}
% \end{equation}
The LSTM training procedure  learns the weights $\theta_S$ by minimizing the loss between the predicted   and the actual average throughput, $\hat{Y}^{(i+\mathcal{W})}$ and $\overline{Y}^{(i+\mathcal{W})}$ (Line \ref{algo_learn_LSTM}), respectively, i.e.,
\begin{equation}\label{eq:weight_Tune}
    \mathbf{\theta_{S}}  = \argmin{\theta_S} loss(\hat{Y}_S^{(i+\mathcal{W})}, \overline{Y}_S^{(i+\mathcal{W})}), \hspace{1cm} \forall i\in \tau_\mathcal{S}.
\end{equation}
Here, the predicted throughput $\hat{Y}^{(i+\mathcal{W})}$ is obtained as,
\begin{align}\vspace*{-0.5cm}\label{eq:ypred}
     \hat{Y}_S^{(i+\mathcal{W})} =  \mathcal{F_S}((\mathbf{\psi}_{X,S}^i,\psi_{Y,S}^i),\overline{Y}_S^{(i+\mathcal{W})}, \mathbf{\theta_S}),
\end{align}
and the actual average throughput $\overline{Y}_S^{(i+\mathcal{W})}$ is given by:
%Here $\overline{Y}^{(i+T)}$ is the actual average throughput over the future $T$ seconds given by,
\begin{equation}\label{eq:thpt_avg}
   \overline{Y}_S^{(i+\mathcal{W})} = \frac{1}{\mathcal{W}}\sum_{j=i}^{i+\mathcal{W}}Y_{S}^{(j)}.
\end{equation}
Here, learning the predictive function  $\mathcal{F_S}$ is equivalent to learning the weights $\theta_S$ for the model $M_S$. Once the source domain trained LSTM model $M_S$ is available, the next step is the retraining of the layers of the neural network for the target domain $D_T$.
\paragraph{\textbf{The Target Domain}} The target domain dataset $\mathcal{D}_T$ is first preprocessed to extract the features (\S\ref{sec:featurespace}) which are common with the source domain dataset $\mathcal{D}_s$ (Line \ref{algo_preprocess}). After the feature selection, the target domain dataset $\mathcal{D}_T$ is normalized using Max-Min normalization techniques (Line \ref{algo_normalize}). The normalized dataset $\mathcal{D}_{norm}$ is a collection of data points $\{\mathbf{X}_T^{i} , Y_T^{i}\}$, over an observation time $\tau_T$.  $\mathbf{X}_T^{i}$ correspond to only those network parameter and location related features as $\mathbf{X}_S$.  Matrices $(\mathbf{\psi}_{X,T}^i, \psi_{Y,T}^i)$ are then created in the same way as Equation~(\ref{eq:createsample}) and Equation~(\ref{eq:createsample_thpt}) to override the time-series format (Line \ref{algo_creattargetsample}). Finally, we fine tune $\mathcal{F_S}$ using the dataset $\mathcal{D}_T$ in the target domain such that the error in predicting the throughput in the target domain is minimized (Line \ref{algo_learn_LSTM_target}) i.e.,
\begin{equation}
         \mathbf{\Delta\theta^{*}} = \argmin{\Delta\theta} \mathrm{loss}(\hat{Y_T}^{(i+\mathcal{W})}, \overline{Y_T}^{(i+\mathcal{W})}), \hspace{0.5cm} \forall i\in \tau_T,
\end{equation}
where $\hat{Y}_T^{(i+\mathcal{W})}$ is the predicted average throughput over a finite future time window of length $T$ in the target domain. 
\begin{align}\vspace*{-0.5cm}\label{eq:ypredtarget}
     \hat{Y}_T^{(i+\mathcal{W})} =  \mathcal{F}_T((\mathbf{\psi_{X,T}^i},\psi_{Y,\mathcal{\mathcal{W}}}^i),\overline{Y_T}^{(i+\mathcal{W})}, \mathbf{\theta_S}+\Delta\theta),
\end{align}
and  $\overline{Y_T}^{(i+\mathcal{W})}$ is the actual average throughput over $\mathcal{W}$ in the target domain.

 \textbf{Fine Tuning:} In this work, the new weights $\theta_T$ have been obtained from $\theta_S$ as $\theta_T=\theta_S+\Delta\theta$ by retraining the source domain LSTM model using the following inductive bias transfer techniques - (i) weight transfer from source domain ~\cite{yang2015learning}, (ii) partial retraining of layers~\cite{xuhong2018explicit}, (iii) retraining all layers ~\cite{yosinski2014transferable}.
%  \begin{enumerate}
%      \item \textbf{Weight Transfer}: In this technique, the pre-trained model $M_S$ and the predictive function learnt are directly evaluated on the target domain dataset $\mathcal{D}_T$~\cite{yang2015learning}.
%      \item \textbf{Fine Tuning Partial Layers}: This technique  freezes the weights of the initial layers and retrains the last few layers of the model $M_S$ with the target domain data~\cite{xuhong2018explicit}. It requires minimal training time and data. 
%      \item \textbf{Fine Tuning All Layers}: This technique involves fine-tuning the weights of all layers of $M_S$ using the target domain dataset~\cite{yosinski2014transferable}. The training time is still much less than full-fledged model training.
%  \end{enumerate}
 Only the last layer of the LSTM model $M_S$ has been retrained for partial retraining. Further, details  are provided in \S\ref{section:evaluation}.
\subsection{The Buffer Length Decision Engine}\label{sec:buff_length_dec_engine}
 %In this section, we describe the design aspects of the buffer and the bitrate decision engines. \\
 \indent In addition to the throughput prediction engine, the Buffer Length Decision engine also runs at the beginning of each time slot. Once the network throughput is predicted, it is used to predict the optimal playback buffer length so that the energy consumption is minimized. However, the relationship between the predicted throughput and the buffer length is not straightforward and is not analytically tractable. So, we employ an A3C \ac{RL} based deep neural network to determine the optimal buffer length.\\ 
 %As mentioned earlier, we run \ac{RL} algorithm to predict the buffer length.  %\basabdatta{The buffer length is tuned to the wireless network throughput. So, it is imperative that the buffer length be tuned accurately to the predicted average throughput, any aberrations from which will initiate a gradual degradation in the opportunities for energy savings and QoE improvement.} {\textcolor{red}{The previous two lines highlighted in blue may be removed.}} However, the relationship between the predicted throughput and the buffer length is not straightforward and hence is not analytically tractable. This encourages us to employ a deep-RL based neural network to reach the buffer length decision.\\
%\indent Since we use an RL algorithm operates on an environment \ti{E}, takes a state \ti{S} as input, responds with an action \ti{A}, and receives reward \ti{R}. 
\indent The components of the \ac{RL} algorithm corresponding to the buffer decision engine are as follows:\\
% \begin{itemize}
\noindent \textbf{(i) Environment}  \ti{E} - video player.\\
\noindent \textbf{(ii) Input state at timeslot \mq{t}} $\Omega_t$ =  ($\hat{Y}_{{t}}$, $B^{av}_t$, ${\mathcal{X}}$), where $\hat{Y}_{{t}}$ is the  average predicted cellular network throughput, $B^{av}_t$ is the current playback buffer capacity available in seconds,
and $\mathcal{X}$ is the set of possible changes in buffer length. As the current buffer length is $B^{av}_t$, the next buffer length can be predicted to be $\hat{B}_{t+\mathcal{W}}= B^{av}_t+x$ where $x\in\mathcal{X}:=\{-2,-1,0,+1,+2\}$. $x=1$ implies an increase in buffer length by a single chunk, each chunk is of eight seconds.\\
\noindent \textbf{(iii) Action} $A_t$ - Decisions on the increase or decrease of buffer length at timeslot $t$. \\
\noindent \textbf{(iv) Reward} - The reward function is designed in such a manner that the QoE maximized and the energy consumption is minimized as shown in (\ref{eq:reward_bufflen}).
%as a linear weighted function of energy savings with respect to a baseline ABR algorithm and the QoE score:
    \begin{equation}\label{eq:reward_bufflen}
    \Xi_{\mathrm{bufflen}} = w_1.QoE + w_2. (\left|E_{\mathrm{EnDASH-5G}_t}-E_{\mathrm{old}_t}\right|)
    \end{equation}
    %     \begin{equation}\label{eq:reward_bufflen}
    % \Xi_{\mathrm{bufflen}} =QoE - \alpha E_{\mathrm{EnDASH-5G}_t} 
    % \end{equation}
    
% \end{itemize}
\indent The first term in (\ref{eq:reward_bufflen}) is the \textbf{\ac{QoE}} metric  \cite{Yin2015}:
\begin{equation}
    \begin{split}
        \label{eq:QoE}
        \mathrm{QoE} = & \frac{1}{N}\sum_{i=1}^N q(R_i) - \frac{1}{N}\mu\sum_{i=1}^N \delta_i \\
        & - \frac{1}{N-1}\sum_{i=1}^{N-1}\left|q(R_{i+1})-q(R_i)\right|
    \end{split}
\end{equation}
The \textbf{QoE} metric is defined for a video with $N$ chunks. $R_i$ is the bitrate of chunk `$i$' in    bits per second, and $q(R_i)$ maps the bitrate to a quantity that represents the quality perceived by the user. We have taken $q(R_i) = R_i$ \cite{mao2017neural}. $\delta_i$ is the rebuffering time in seconds, incurred while downloading the chunk `$i$' at rate $R_i$. $\mu$ is the degree of penalty associated with $\delta_i$. We have taken $\mu=4.3$ \cite{mao2017neural}. The last term represents the playback smoothness. The QoE increases when the streamed video is smooth, i.e., there is less variation in the bitrate of subsequent chunks. Smoothness is quantified by its inverse, which is the throughput variation having a unit of bits per seconds. QoE, thus, increases with bitrate and reduces with rebuffering time and throughput variability.\\
\indent The second term in (\ref{eq:reward_bufflen}) gives the energy savings with respect to a baseline ABR algorithm. In this work, we have chosen BOLA~\cite{Spiteri2016} as the baseline since its energy consumption is the lowest among existing algorithms (excluding EnDASH-5G, as seen in \S{\ref{sec:EnDASHresults}}). $E_{\mathrm{old}_{t}}$ and $E_{\mathrm{EnDASH-5G}_t}$ represent the energy consumption of BOLA and of EnDASH-5G at time slot \mq{t}, respectively. When the energy consumed by EnDASH-5G is minimized, the difference in the energy consumed by EnDASH-5G and BOLA increases. Thus, the second term in (8) also increases. Consequently, overall we attempt to maximize the reward which is a weighted combination of the QoE and the energy difference. We have introduced the two weights for generalization. For the current work we have taken $w_1=w_2=1$. These may be controlled to handle the tradeoff according to the user's requirement. \\
\indent We choose this energy minimization method in order to compensate for the lack of ground truth on the energy consumption of the ABR streaming algorithms. By using this method, pre-trained models can be loaded into smartphones.
The energy consumed is obtained as follows:  the RRC states are first identified from the download packet capture of a video trace. Next, the dwell time in each RRC state is multiplied by the corresponding power consumption (obtained from the RRC state machine) to get the energy quantities.
%We calculate the energy savings in this manner because the ground truth on energy consumption cannot be obtained directly from the device.\\
%It is obtained as follows. The buffer length decision engine is trained using the power consumption and throughput data traces that we have collected using the Moto G5 phone over the Airtel network. Corresponding to this handset and network combination we have obtained an RRC state machine which gives the power consumptions in the \textit{CONNECTED}, \textit{TAIL}, and \textit{IDLE} RRC states,  as discussed in \S\ref{section:motivation}. From a collected video trace it is possible to identify these state. So, equipped with the RRC state machine it is possible to calculate the energy consumption of any \ac{ABR} streaming algorithm, once the RRC states are identified for any video trace.
%\indent Second term 
\subsection{The Bitrate Decision Engine}
 \label{sec:bitrate_dec_engine}
The third component of EnDASH-5G is the bitrate decision engine which runs within a time slot and is used to predict the bitrate of the next video chunk immediately before it is downloaded.  The buffer length and bitrate decision engines run at two different time scales- buffer length decision engine at the beginning of a time and the bitrate decision engine within a time. So, time slot indices for the variables related to the buffer length decision engine and the bitrate decision engine are denoted by  \mq{t} and \mq{n}, respectively.The bitrate decision engine is also an A3C-RL based deep neural network whose components are as follows:\\
%\begin{itemize}
\noindent\textbf{(i) Environment}  \ti{E} -- video player.\\
\noindent\textbf{(ii) Input state before downloading the chunk \mq{n}}, \\$\Omega_{n}$ = ($\hat{\mathcal{B}}_t$, $y_c(n-1)$, $d_{n-1}$, $l_{n-1}$, $B_{n}$, $\beta_{n}$, $r_{n}$), where $\hat{\mathcal{B}}_t$ is the predicted playback buffer length for current slot \mq{t}, $\tau_c(n-1)$ is the throughput of the last chunk, $d_{n-1}$ is the time taken to download last chunk, $l_{n-1}$ is the bitrate of the last chunk, $B_{n}$ is the available playback buffer length, $\beta_{n}\in\bracket{b_1,b_2,\cdots b_m}$ is the possible size of next video chunk  ($m$ available sizes), $r_{n+1}\in \bracket{R_1, R_2,\cdots, R_6}$ is the possible bitrate levels for next video chunk.\\
\noindent\textbf{(iii) Action} $A_n$ -- Optimal bitrate decision for the next chunk.\\ 
\noindent\textbf{(iv) Reward} -- QoE score obtained from Equation (\ref{eq:QoE}).
\section{Implementation and Training of ML Models}
\label{sec:imp} 
In this section, we explain the implementation details of the three decision engines of  EnDASH-5G.
%uses three decision engines -- (i) the throughput prediction engine, (ii) the buffer length decision engine, and (iii) the bitrate selection engine. \textcolor{black}{Both the buffer length decision engine and bitrate decision engines are trained simultaneously}. The details follow. 
% Based on the earlier Pilot study in \ref{section:pilot} we generate 5G simulation traces of video download under different network scenarios. This \textbf{target domain 5G dataset} along with the \textbf{source domain 4G dataset} (same as that of \cite{Mondal2020}) are being used to develop an LSTM based transfer learning model, to train and test the throughput prediction engine of EnDASH. We have divided the collected dataset into 80-20 ratios for training and testing. Furthermore, to train the RL algorithms of the buffer length and bitrate decision engine, we have used another dataset of 57 DASH-ified videos having a total duration of 45 hours.

% The \textbf{ABR Server} as discussed in the \ref{ssxn:serverClient} uses EnDASH algorithm to determine the optimum video chunk bitrate for streaming the video over HTTP. Next, we compare the performance of EnDASH with other baseline ABR Algorithm.

\subsection{Dataset used for Throughput Prediction}

\subsubsection{Source Domain Datasets}
The {source domain 5G datasets} include three datasets as explained next- 
%(i) the \textbf{Lumos-5G dataset} from \cite{Narayanan2020},  (ii) the \textbf{Irish dataset} from \cite{Raca2020_1}, and (iii) \textbf{MN-Wild dataset} from \cite{Narayanan2021}. }
\begin{itemize}
    \item \textbf{Lumos-5G dataset}\cite{Narayanan2020} - The Lumos-5G dataset from \cite{Narayanan2020} contains a measurement study of commercial 5G mmWave services in Minneapolis, USA, and has recorded the downlink throughput as perceived by applications running on \ac{UE}. It contains information on the \ac{UE}'s geographical coordinates, moving speed, compass directions, downlink throughput (reported using \textit{iperf 3.7}), radio type (4G/5G), signal strength (LTE -- RSRP, RSRQ, RSSI \& 5G -- SSRSRP, SSRSRQ, SSRSSI), handover events, etc. With Verizon's 5G Ultra Wideband network, the measurement study has been conducted over a period of $6$ months, using $4$ Samsung Galaxy S10 5G phones.
    \item \textbf{Irish dataset} - In \cite{Raca2020_1}, the authors' have collected 5G trace datasets from a major \textbf{Irish} mobile operator. They have considered two application patterns (video streaming and file download). The dataset consists of (1) channel-related metrics such as signal strength (RSRP, RSRQ, SNR, CQI), neighboring cell RSRP, RSRQ, (2) context-related metrics (e.g., GPS of the device, device velocity), (3) cell-related metrics such as eNBs ID, and (4) throughput information(both uplink and downlink). The dataset consists of $83$ traces, with a total duration of $3142$ minutes, using a Samsung S10 5G Android device.
    \item \textbf{MN-Wild} dataset - In \cite{Narayanan2021}, authors' have carried out an in-depth measurement study of the performance, power consumption, and application QoE of commercial 5G networks in the wild. They have examined different 5G carriers (Verizon and T-Mobile), deployment schemes (Non-Standalone, NSA vs. Standalone, SA), radio bands (mmWave and sub-6-GHz), Radio Resource Control state transitions for power modeling, mobility patterns (stationary, walking, driving), client devices (such as Samsung Galaxy S20 Ultra 5G (S20U) and Samsung Galaxy S10 5G (S10)), and upper-layer applications (file download, video streaming, and web browsing). The measurement studies were conducted in two US cities (Minneapolis, MN, and Ann Arbor, MI), where both carriers have deployed 5G services. Of the datasets belonging to the two cities, we have found that the dataset corresponding to the Minneapolis city (MN) contains throughput and other important UE information, such as the speed of the UE, which is not available for the Ann Arbor (MI) dataset. Therefore, in this paper, we have selected the Minneapolis dataset and referred to it as MN-Wild.
\end{itemize}

% The distance of UE from the \ac{BS} is not available from the \texttt{NetMonitorLite} app. The  app records the Mobile Country Code (MCC), Mobile Network Code (MNC), Location Area Code (LAC) and Cell Id (CID) of the associated base station. Using these metrics the geographical coordinates of the \ac{BS} were obtained from \textbf{OpenCellId}\footnote{OpenCelliD: \url{https://opencellid.org/} (Accessed: \today)}. The distance of \ac{UE} from the \ac{BS} was calculated from the \ac{UE} and \ac{BS} coordinates.
 \subsubsection{Target Domain Datasets}

 As mentioned earlier, the\textbf{ target domain dataset}  is the \textbf{simulated 5G network}. The corresponding dataset is the  synthetic throughput data collected from the $\mathtt{ns3-mmWave}$ simulation outlined in \S\ref{section:pilot}. The target domain dataset has 23778 datapoints, captured at the granularity of 0.44 seconds on an average.\footnote{In ns-3 a datapoint is recorded at each event trigger. Hence, the granularity of the datapoints is not constant.}
 \subsubsection{\textbf{Identifying the feature space}}\label{sec:featurespace} As identified in \cite{Mondal2020} the cellular network throughput depends on several parameters -  the \ac{UE} parameters such as user speed and distance from \ac{BS}, as well as the network related parameters such as \ac{RSSI}, \ac{RSRP}, \ac{RSRQ}, number of handovers, technology of associated and neighbouring \acp{BS}, \ac{MCS}, data state (CONNECTED or IDLE). In this work, we have first identified a standard set of features which are common to both the source and the target domain datasets. These are, user speed,  distance from \ac{BS}, \ac{RSSI}, \ac{RSRP}, \ac{RSRQ}, number of handovers, data state.

%  \argha{\textcolor{blue}{
 %We need a common feature space for the source and the target domain dataset to execute throughput prediction using transfer learning. Correspondingly, the features used to enable transfer learning from the 5G public dataset and the \textbf{simulated 5G mmWave dataset} are -(1) \textbf{Radio Channel metrics}: \ac{RSSI}, Handover (2) \textbf{UE metrics}: UE speed, Downlink Throughput.
\indent We train the LSTM model of the throughput decision engine with the public-5G datasets~\cite{Narayanan2020, Narayanan2021, Raca2020_1} with a train-validation split of $80\%$-$20\%$. The trained model is then retrained by means of Transfer Learning using the ns-3 simulated 5G dataset. For the retraining we have used a train-validation-test split of $72\%$-$8\%$-$20\%$. This tunes the hyperparameters of the LSTM model. We have provided the prediction $R^2$ score on the test set in our evaluation (Table~\ref{tab:pred_dataset}). The trained model is then deployed in the ns-3 simulation setup for testing with new throughput data generated in runtime.
\subsection{Emulation Platform for \ac{RL} Training}
     In the training phase, the \ac{RL} agent of the buffer and the bitrate decision engines of EnDASH-5G should ideally be trained using real video downloads at actual video streaming clients. 
  %This requires that an entire video chunk be downloaded for a single training data point. 
Emulating a standard video streaming session demands   that for each training data point, all the video chunks are downloaded over a web browser before training can start. This unnecessarily lengthens the training time. As downloads over cellular networks can be pretty slow due to network fluctuations, the training time can become longer. Hence, to save training time, we train EnDASH-5G and its competing ABR algorithms using a python-based video emulation setup that closely mimics a real video client application as in~\cite{mao2017neural, Sengupta2018, Mondal2020}. Our source code is publicly available in GitHub~\footnote{\url{https://github.com/arghasen10/endash/tree/main/rl_train} (Accessed:\today)} \\
 \indent The emulator maintains its own representation of a real client's playback buffer compatible with the Mahimahi~\cite{Netravali2015} network emulation tool. At the beginning of a time slot, the emulator first predicts the average throughput and then the playback buffer length for the slot. When a chunk is to be downloaded within the slot, the emulator first assigns a download time to the chunk based on - a) its bitrate and b) the internal network throughput derived from previous chunks. It then depletes the playback buffer by the current chunk's download time to emulate the playback buffer drainage during an ongoing chunk download. Finally, it adds the playback duration of the downloaded chunk to the playback buffer. The emulator sleeps temporarily once the playback buffer is full. Rebuffering occurs if the playback buffer is completely drained before the next chunk download. The emulator keeps track of the rebuffering event and the rebuffering time. At the end of each slot and chunk download, the emulator prepares the state $\Omega_t$ and $\Omega_n$ for the \ac{RL} training algorithm of the buffer and bitrate decision engines, respectively.
 \subsection{The \ac{RL} Training Algorithm}
Both the buffer length and bitrate decision engines of EnDASH-5G are trained using the state-of-the-art actor-critic method, which involves training two neural networks \cite{mao2017neural}: an actor-network and a critic network. A3C is a policy gradient \ac{RL} algorithm that estimates the gradient of the total reward from the trajectories of the execution followed by a policy. The action is chosen based on the policy by the actor-network, while the critic network estimates the advantage of selecting the action by returning a value function. Both networks update their weights in each time step. \\
 \indent Each parameter of the state $\Omega$ of both the buffer length and bitrate decision engines are passed on to their respective actor and critic networks to enable learning of optimal buffer lengths and bitrates by the corresponding \ac{RL} modules~\cite{mao2017neural}.
%\\
%\indent At the end of each time slot, the  \ac{RL} agent of the buffer length decision engine takes the State Input $\psi_T$ as discussed in  \S\ref{sec:buff_length_dec_engine}. While once a video chunk is downloaded, the \ac{RL} agent  of the bitrate decision engine takes the State $\psi_n$ as input. $\psi_n$ is defined in \S\ref{sec:bitrate_dec_engine}. \\
% Once the respective buffer length and bitrate \ac{\ac{RL}} agents receive the states $\psi_T$ and $\psi_n$, they take the actions $A_T$ and $A_n$ corresponding to these states. . 
 %The actions are taken by the two engines based on  policies which are defined as a probability distributions over the actions, i.e., $\pi(\psi_T,A_T)\longrightarrow [0,1]$ and $\pi(\psi_n,A_n)\longrightarrow [0,1]$. Both the decision engines learn the policy using some neural networks. After each action is taken, the simulated environment provides the corresponding reward to the learning agent, whose objective is to maximize the reward. \\
 %\textbf{Parallel Training:} 
 As in \cite{mao2017neural}, we initiate multiple learning agents to accelerate the training. By default, there are 16 agents, each of which is designed to encounter a different set of input parameters, e.g., network traces. The learning agents continuously report their individual tuples of (state, action, reward)  to a central model, which aggregates them to generate a single model.
 
\subsection{Implementation of the \ac{RL} Modules}
The \textit{bitrate selection engine} of EnDASH-5G is adopted from the  Pensieve algorithm \cite{mao2017neural}, which also uses A3C to learn optimal bitrates. We have used a pre-trained model provided by Pensieve, albeit with different input states. The \textit{buffer length decision engine} has been implemented in TensorFlow. The engine passes five previous buffer length values to a 1D convolution layer (CNN) with 128 filters, each of size 4 with stride 1. The remaining inputs, i.e., predicted link throughput and currently available buffer length, are passed onto another 1D-CNN having the same shape. Results obtained from these layers are  combined into a hidden layer having 128 neurons and use the softmax function. The critic network uses the same inputs with the same neural network, but its final output is a linear neuron. During the algorithm's training, we set the discount factor to $0.9$, i.e., current actions can influence by $100$ future steps. The learning rates for the actor and the critic networks are $0.0001$ and $0.001$, respectively. Over iterations, the entropy factor is gradually decreased from $1$ to $0.1$.\\
\indent An important point to note is that the buffer length and bitrate decision engine does not focus on hyperparameter tuning. We have used the hyperparameters from~\cite{mao2017neural}.

%\indent In the next section, we discuss how the proposed EnDASH-5G system performs in terms of energy savings and QoE with respect to other state-od-the-art \ac{ABR} algorithms.
\subsection{Simulation Environment for EnDASH-5G}
We have implemented EnDASH-5G ABR Controller over the ns-3 simulation framework outlined in the description of the Video streaming server in \S\ref{ssxn:serverClient}.The models corresponding to the three decision engines, pre-trained on the training dataset, have been ported to the designed ABR Controller. In runtime, the RF decision engines predict the buffer length and video bitrate based on the playback buffer information and the pretrained throughput predictor predicts the network throughput based on the runtime throughput traces. 
%Next, we discuss the results obtained from this thorough simulation. 
	\section{Results}\label{section:evaluation}
\begin{figure*}[t]%
	\centering
	\subfigure[Energy Consumption]{%
		\label{fig:EnDASH_en}%
		\includegraphics[width=0.3\textwidth]{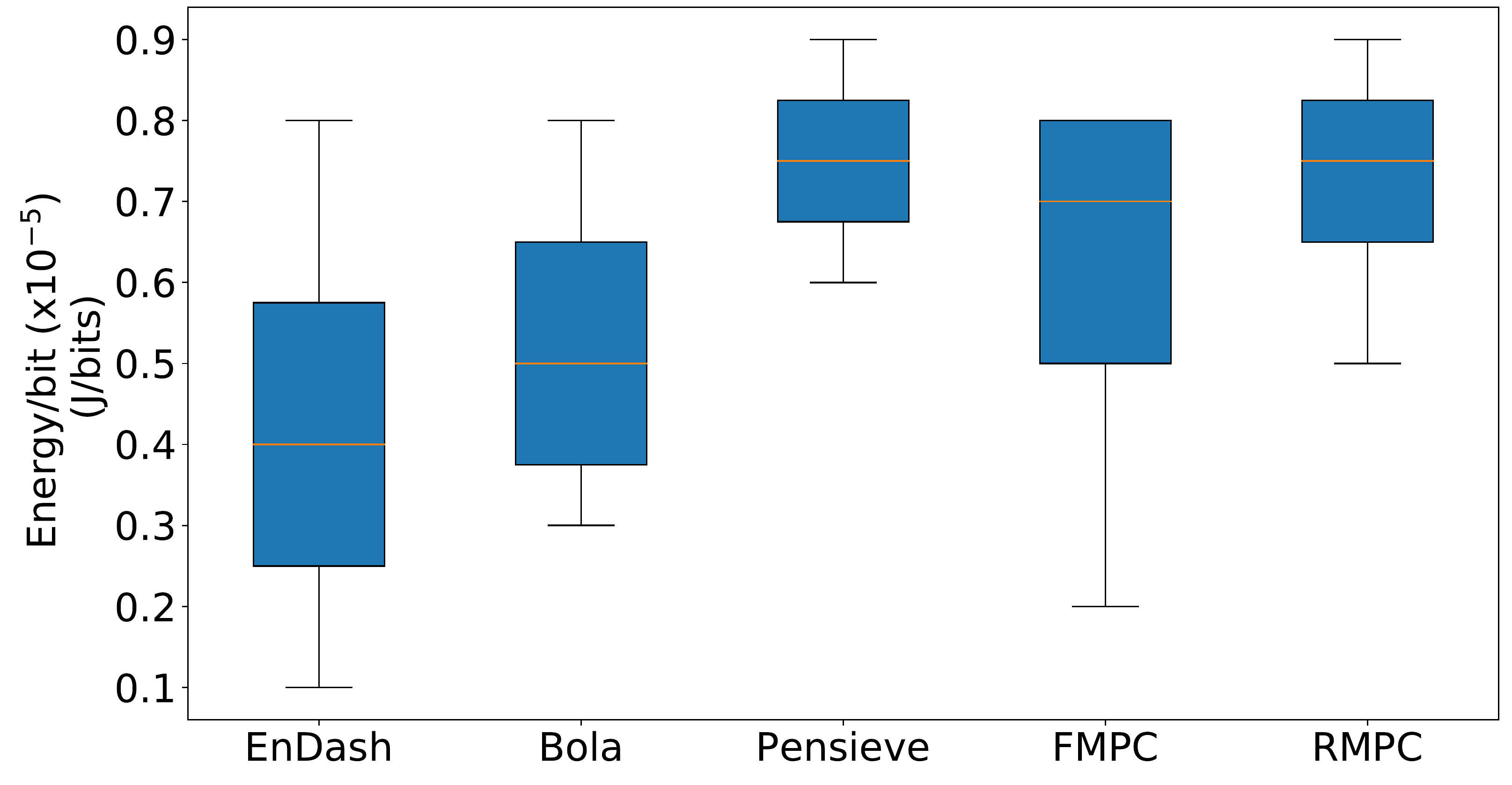}}\hfil%
	\subfigure[QoE score (equation~\ref{eq:QoE})]{%
		\label{fig:EnDASH_QoE}%
		\includegraphics[width=0.3\textwidth]{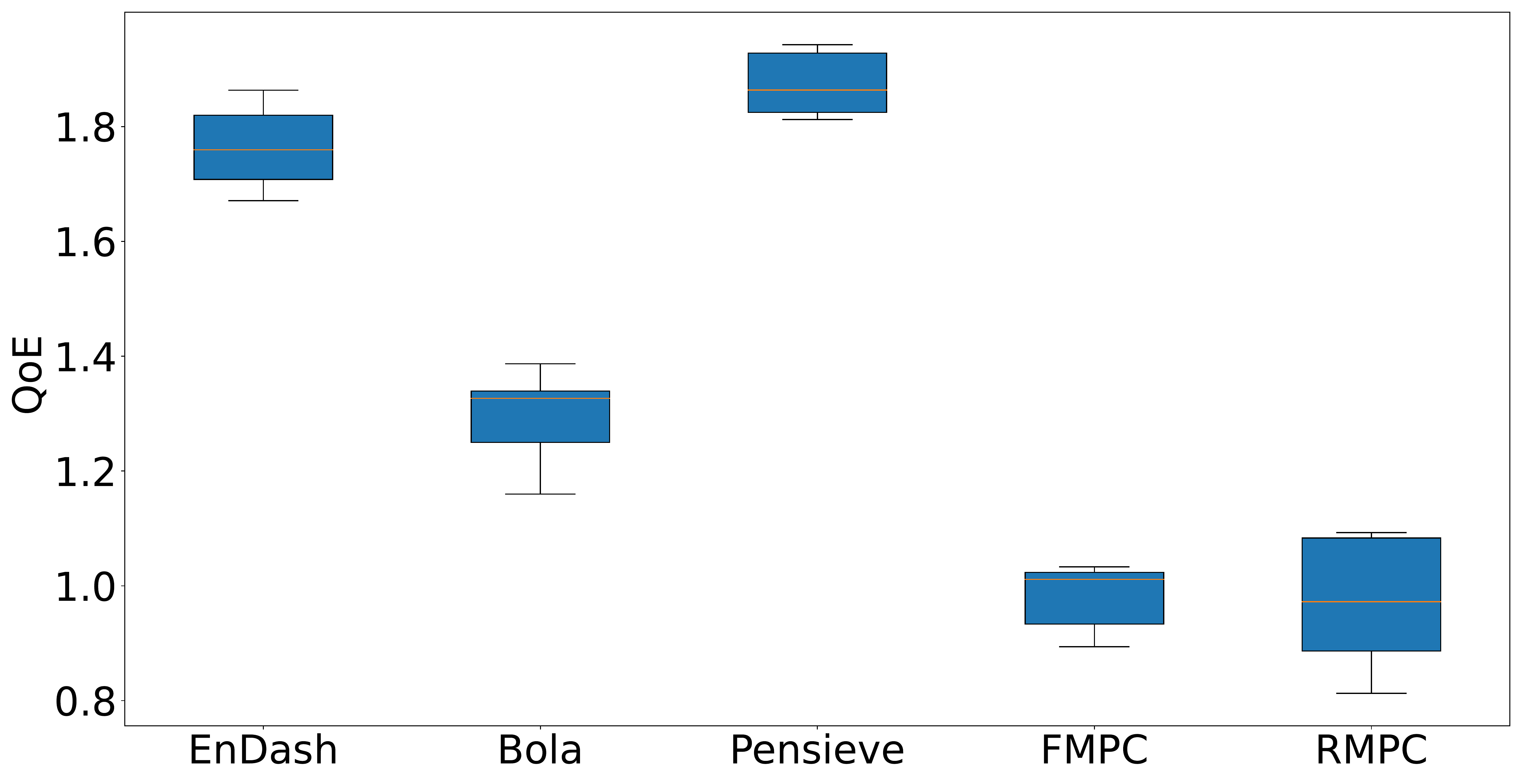}}%
% 	 \subfigure[\textcolor{blue}{Spider chart across QoE and it's components}]{%
	 \subfigure[Scaled Spider chart across QoE components]{%
	 	\label{fig:spider}%
	 	\includegraphics[width=0.3\textwidth]{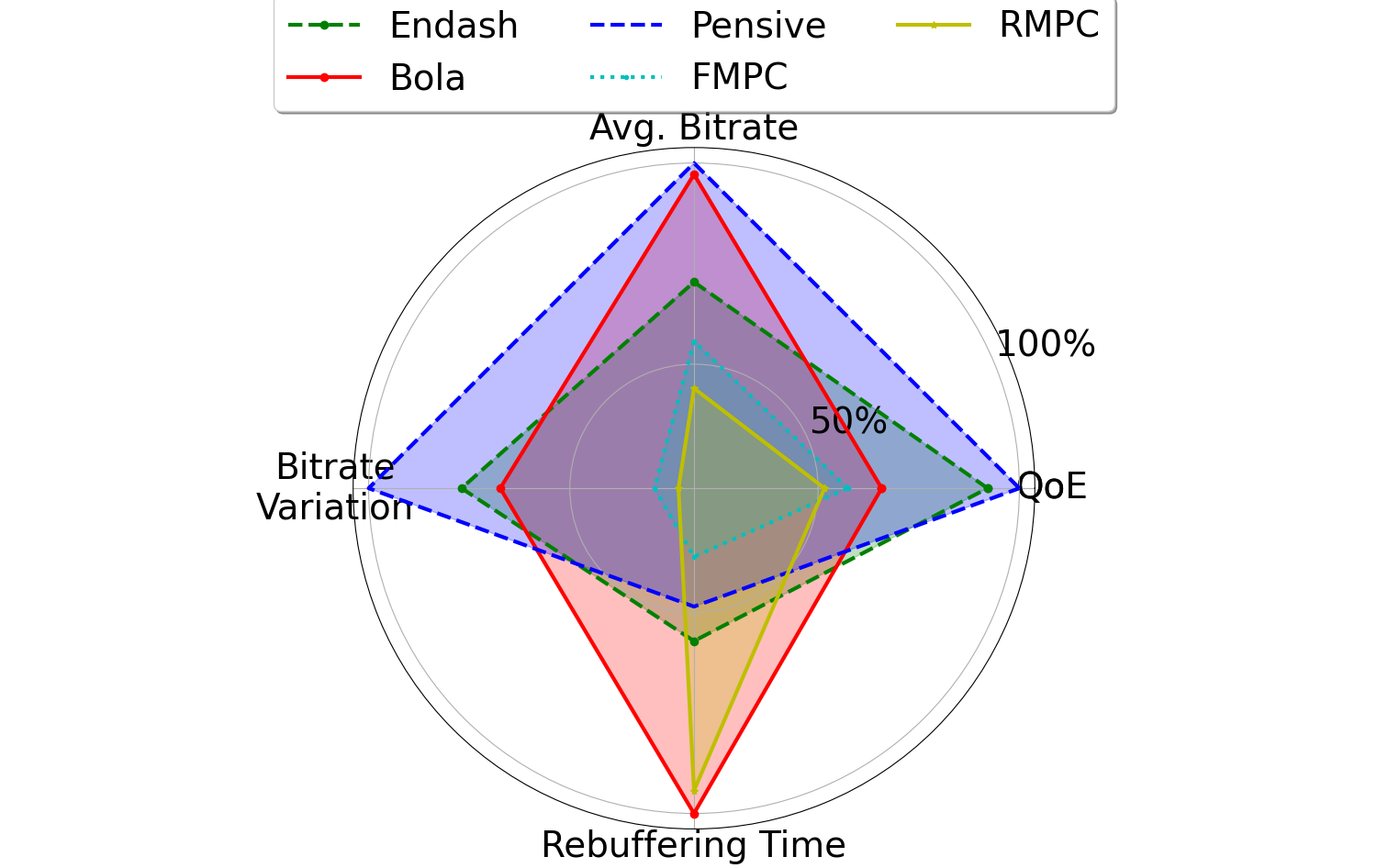}}
	\hspace{0.1cm}
	\caption{Performance comparison of EnDASH-5G with baseline \ac{ABR} streaming algorithms, BOLA \cite{Spiteri2016}, Pensieve \cite{mao2017neural}, \ac{FMPC} \cite{Yin2015}, \ac{RMPC} \cite{Yin2015}. The 100\% in Fig. \ref{fig:spider} correspond to the maximum value of the respective parameter.}
	\label{fig:EnDASH_vs_others}
\end{figure*}
%This section summarizes the observations and results obtained from the thorough simulation of the EnDASH-5G framework. 
We first discuss the overall energy and QoE performance of EnDASH-5G and then dig into its individual modules. 
\subsection{EnDASH-5G versus Baseline ABR algorithms}\label{sec:EnDASHresults}
In this work, we have considered four state-of-the-art ABR schemes as baselines and compared their performances with EnDash-5G. These algorithms include - (i) BOLA~\cite{Spiteri2016} - a \textbf{buffer based} algorithm that tunes the playback buffer length, (ii) \ac{FMPC} \& \ac{RMPC}~\cite{Yin2015} - \textbf{control theoretic} schemes that maximize QoE based on upcoming chunk sizes and predicted future throughput, and (iii) Pensieve~\cite{mao2017neural} - a \textbf{learning-based} approach that use deep neural network decision engine to decide the optimal bitrate.

\begin{figure*}[t]%
	\centering
	%\vspace*{-0.5cm}
	%\hspace{0.1cm}
	\subfigure[Avg. Bitrates]{%
		\label{fig:avg_bitrate}%
		\includegraphics[width=0.28\textwidth]{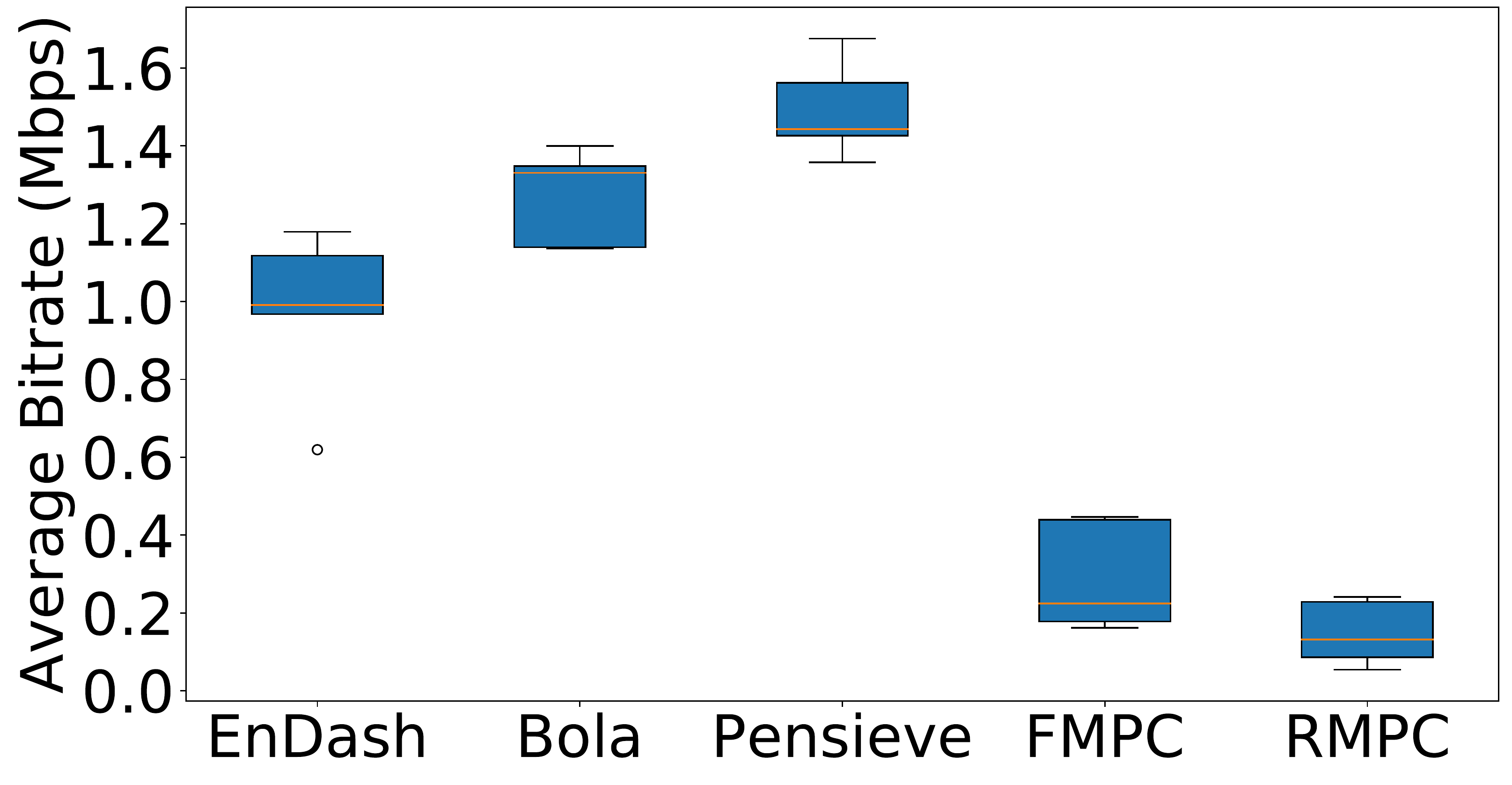}}\hfil%
	\subfigure[Bitrate Variation]{%
		\label{fig:smooth}%
		\includegraphics[width=0.30\textwidth]{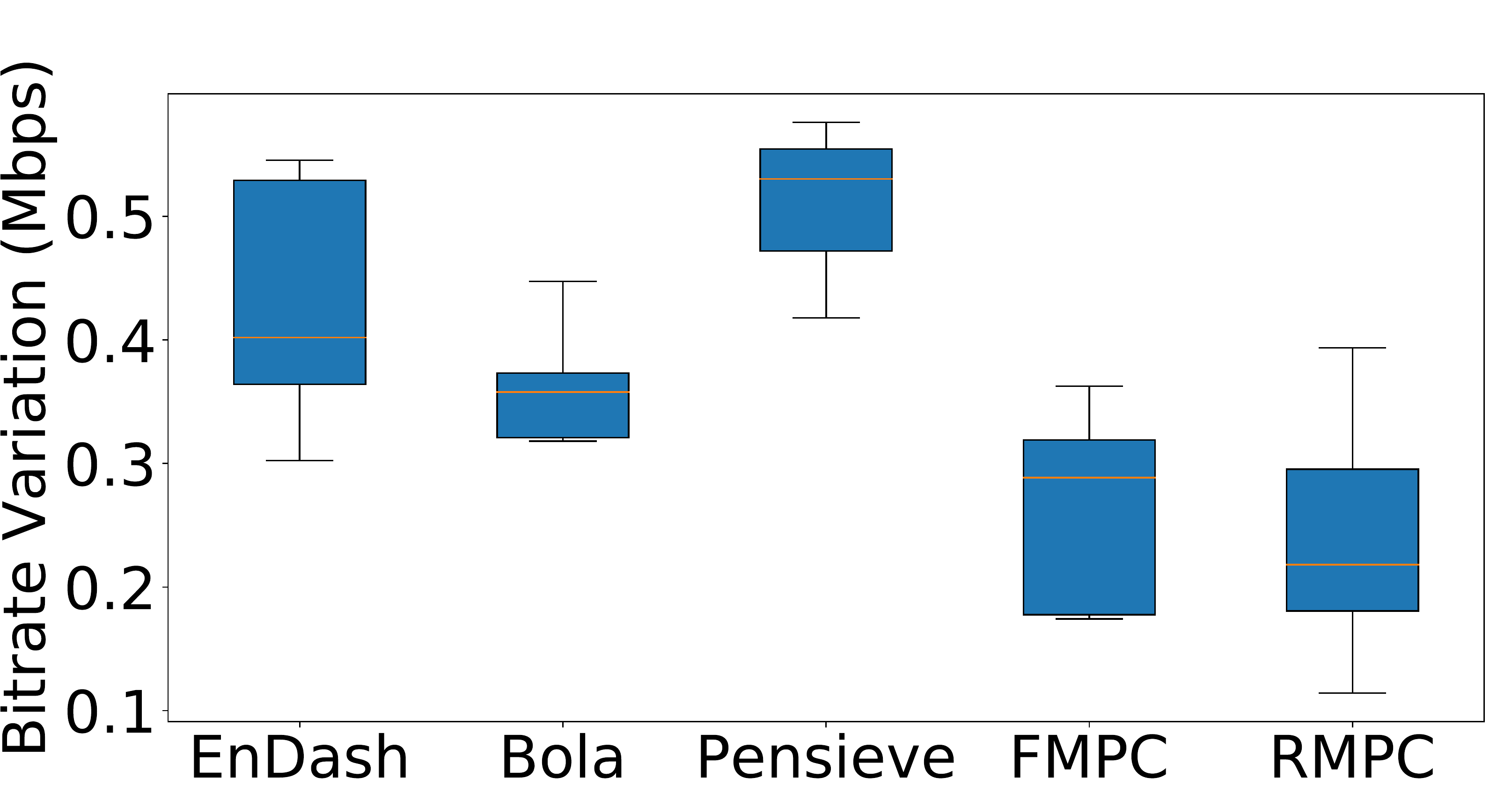}}\hfil%
	\subfigure[Rebuffer Time and average number of rebuffer events]{%
		\label{fig:rebuffer_time}%
		\includegraphics[width=0.30\textwidth]{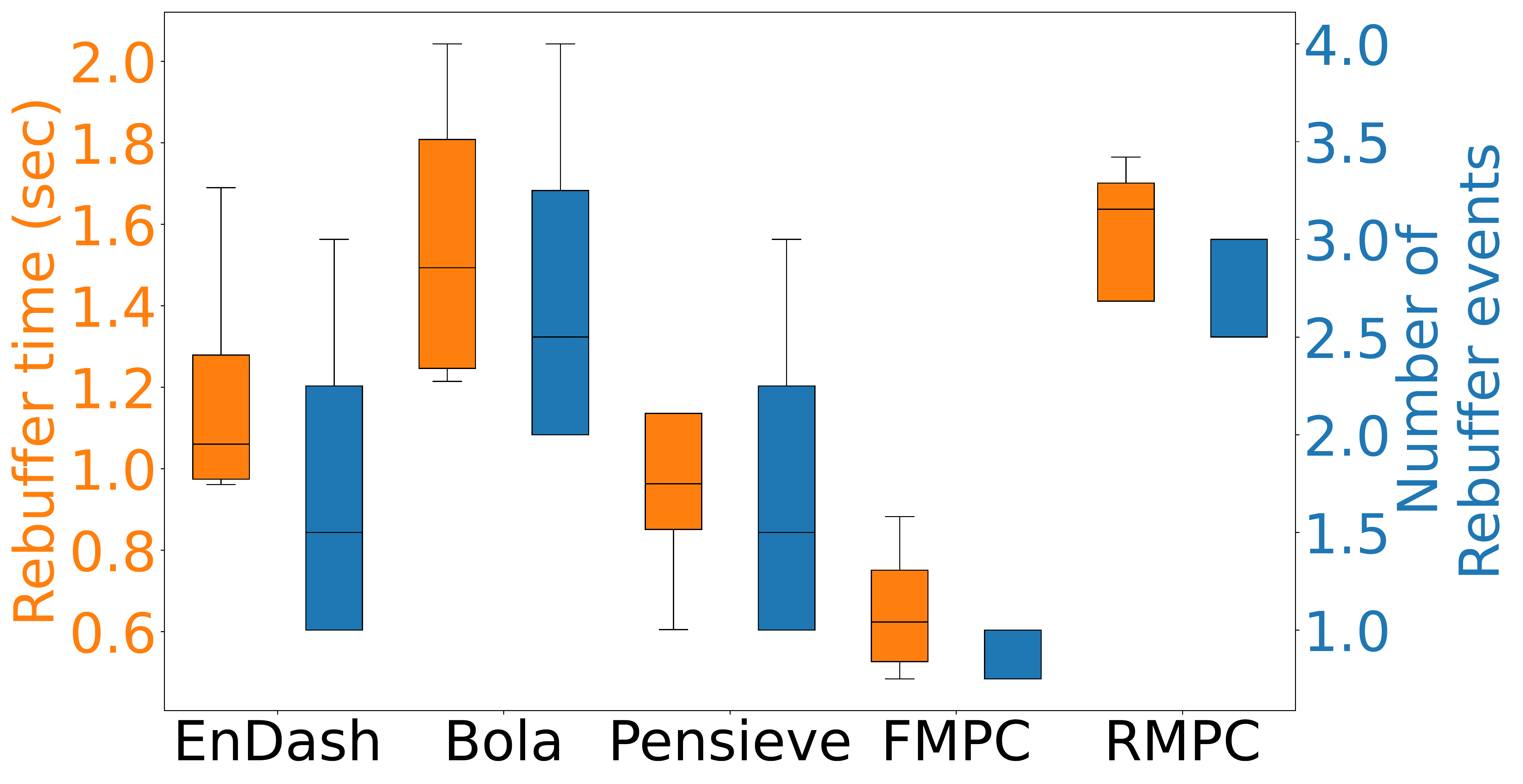}}%
	\caption{Comparison of different components of \ac{QoE} score (average bitrate, bitrate variation, rebuffering time) of EnDASH-5G with baseline \ac{ABR} streaming algorithms, BOLA \cite{Spiteri2016}, Pensieve \cite{mao2017neural}, \ac{FMPC} \cite{Yin2015}, \ac{RMPC} \cite{Yin2015}}\vspace*{-0.5cm}
	\label{fig:indi_QoE}
\end{figure*}

Fig. \ref{fig:EnDASH_vs_others} compares the average energy consumption and the average \ac{QoE} score of EnDASH-5G with the aforementioned baseline \ac{ABR} algorithms, over the 27 different scenarios outlined in Table \ref{tab:sim_scenarios}.  Pensieve~\cite{mao2017neural} has been reported to generate optimal chunk bitrates and hence, delivers a high \ac{QoE} to users. Fig. \ref{fig:EnDASH_en} and \ref{fig:EnDASH_QoE} show that while Pensieve indeed delivers the highest QoE among the different ABR streaming algorithms, it is quite expensive in terms of energy expenditure. EnDASH-5G reduces the mean energy consumption of Pensieve by nearly 30.5\% in 5G-mmWave networks. This reduction, however, comes at the cost of QoE as the QoE of EnDASH-5G is 4.17\% less than Pensieve's as seen in Fig. \ref{fig:EnDASH_QoE}. The bitrate decision algorithm of Pensieve is oblivious to the underlying channel condition. As a result, if it decides upon a higher bitrate during poor link conditions, the algorithm tries  to download chunks at the high bitrate only, thereby getting stuck in the high power CONNECTED state of the RRC state machine (Fig. 1).  In a similar situation, when EnDASH-5G anticipates the increased energy consumption, it immediately switches to a lower bitrate -- commensurate with the link condition. Thus, EnDASH-5G achieves the reduced energy consumption with a minor compromise of \ac{QoE}.% as shown in Figs. \ref{fig:EnDASH_en} and \ref{fig:EnDASH_QoE}
%This leads to increased energy consumption, as is reflected in Fig. 7a.. As a result, the playback buffer length increases during high throughput conditions, thereby facilitating the download of a higher number of chunks in a slot. This consecutive fetching of chunks reduces the dwell time in the RRC INACTIVE state, and hence, the energy consumption as seen in Fig. \ref{fig:EnDASH_en}. 

%The energy savings, however, comes at the cost of minorly sacrificing the \ac{QoE} concerning Pensieve. Fig.~\ref{fig:EnDASH_QoE} shows that the QoE score of EnDASH-5G is $\frac{7}{8}^\mathrm{th}$ of that of Pensieve. 
Thus, it may be inferred that EnDASH-5G trades off the energy usage with the delivered QoE of ABR streaming algorithms. In the following subsection, we review the individual components of the QoE metric to explore further.

\subsection{QoE Performance Analysis}
The individual components of the \ac{QoE} metric corresponding to Fig. \ref{fig:EnDASH_QoE} have been outlined in eq. (\ref{eq:QoE}), and explained in \S\ref{sec:buff_length_dec_engine}. These  are plotted in Fig. \ref{fig:indi_QoE} and their  spider chart is shown in Fig.~\ref{fig:spider}.  Fig.~\ref{fig:spider} and  Fig. \ref{fig:indi_QoE} also summarize the \ac{QoE} behaviour across all the 27 scenarios of Table~\ref{tab:sim_scenarios}. It is observed from Fig.~\ref{fig:spider} and Fig.~\ref{fig:avg_bitrate} that the average bitrate of EnDASH-5G is lower than BOLA and Pensieve.  This is because the buffer length in EnDASH-5G is tuned to the average predicted throughput. On the other hand, BOLA predicts bitrates using playback buffer length only, which manifests in the lower bitrate variation (Fig \ref{fig:smooth}) but the high rebuffering time (Fig. \ref{fig:rebuffer_time}). Pensieve uses the previous chunk download rates to predict the future bitrates. As this captures the instantaneous throughput variation, it reduces the rebuffering time but increases the bitrate variation.\\
\indent In contrast, EnDASH-5G tunes the buffer length to the average throughput of the underlying channel. This may lead to lower bitrates if the average throughput predicted at the beginning of any time slot is low. In such cases, the system fails to exploit the instantaneous high throughput events within the timeslot. However, tuning buffer length to throughput yields lower bitrate variation than Pensieve. The number of rebuffering events and rebuffering time of EnDASH-5G is also comparable to Pensieve. This compensates for the reduced bitrates of EnDASH-5G, and in combination with its reduced energy consumption, it emerges as a viable candidate for delivering appreciable QoE to streaming videos. 
 %shows that is relatively high. 
%  Fig. \ref{fig:rebuffer_time} shows that EnDASH-5G gains significantly in terms of rebuffering time. This is primarily because of tuning the chunk downloads to the network throughput variation. Due to the high accuracy in throughput prediction and playback buffer length, EnDASH-5G does not overcommit to higher data rates in the face of poor throughput. This improves the rebuffering scenario. 
\subsection{Energy Performance Analysis}
\begin{figure}[!ht]
    \centering
    \includegraphics[width=0.8\linewidth]{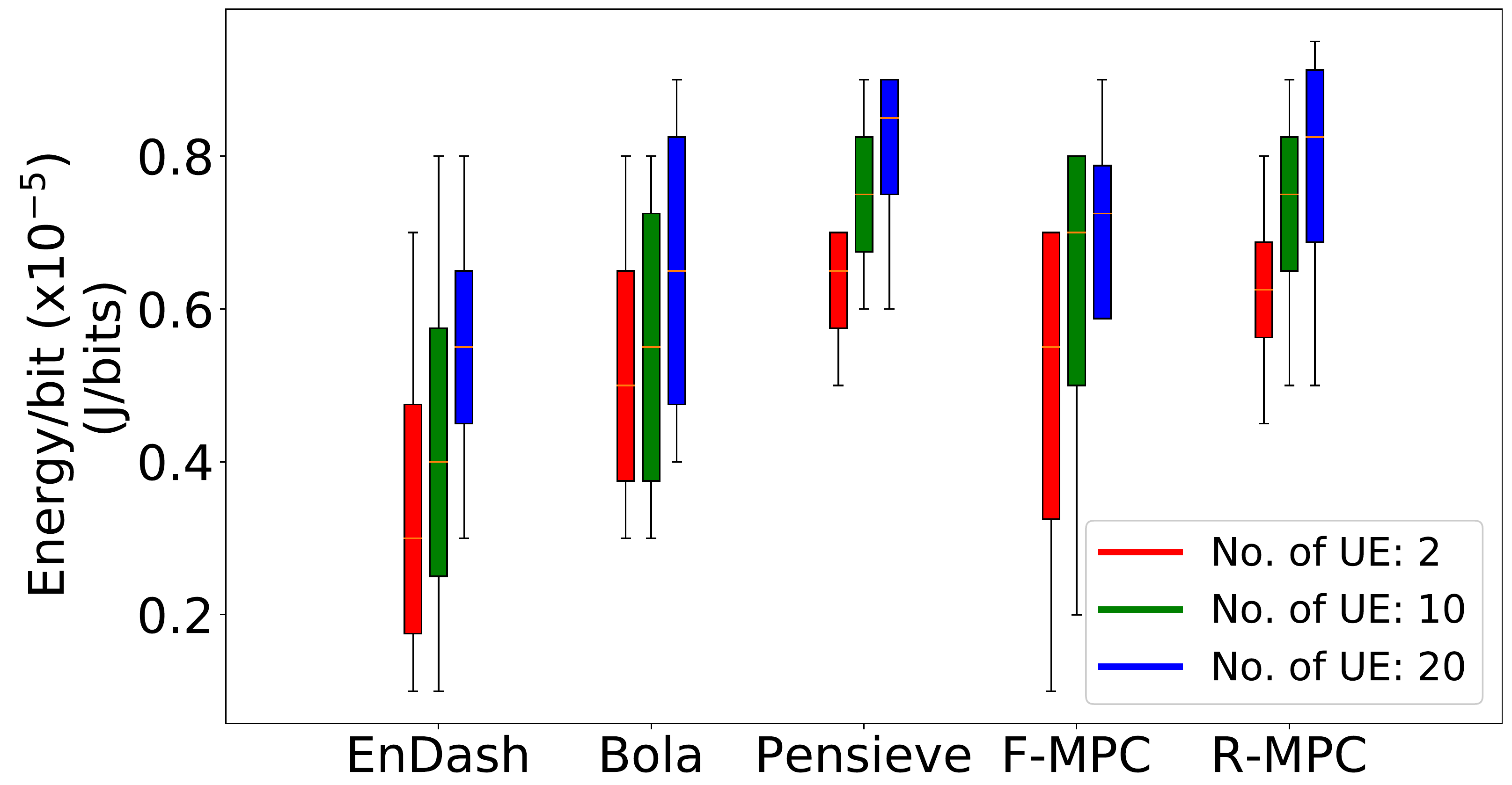}
    \caption{Energy consumption of EnDASH-5G and other baseline algorithms under different network load}
    \label{fig:EnDash_energy_diff_scn}
\end{figure}
%Fig. \ref{fig:EnDash_energy_diff_scn} shows that  EnDASH outperforms every baseline ABR Algorithms in the energy consumption domain. In fact, EnDASH improves the energy consumption by Pensieve by about 27\%. 

Our objective has been to improve the energy consumption even when the network load increases. An increased network load implies fewer radio resources would be available per \ac{UE}, which will decrease  individual user throughput. To evaluate how EnDASH-5G performs under different network loads, we have plotted the energy consumption of EnDASH-5G and other ABR algorithms for a different number of \acp{UE} ( $2, 10, \text{and} \ 20$ users) in the system. Further, for each network load and each \ac{UE} speed, we have varied the number of buildings in the one square kilometre area as $0, 10$, and $20$. Averaging over the network conditions and the \ac{UE} speed for a given network load, we have plotted the energy consumption of the different ABR streaming algorithms under different network loads in Fig. \ref{fig:EnDash_energy_diff_scn}. It is observed that EnDASH-5G keeps the energy consumption significantly lower than other algorithms even at higher network load. 

\begin{figure}[t]%
    \scriptsize
	\centering
	\includegraphics[width=0.8\linewidth]{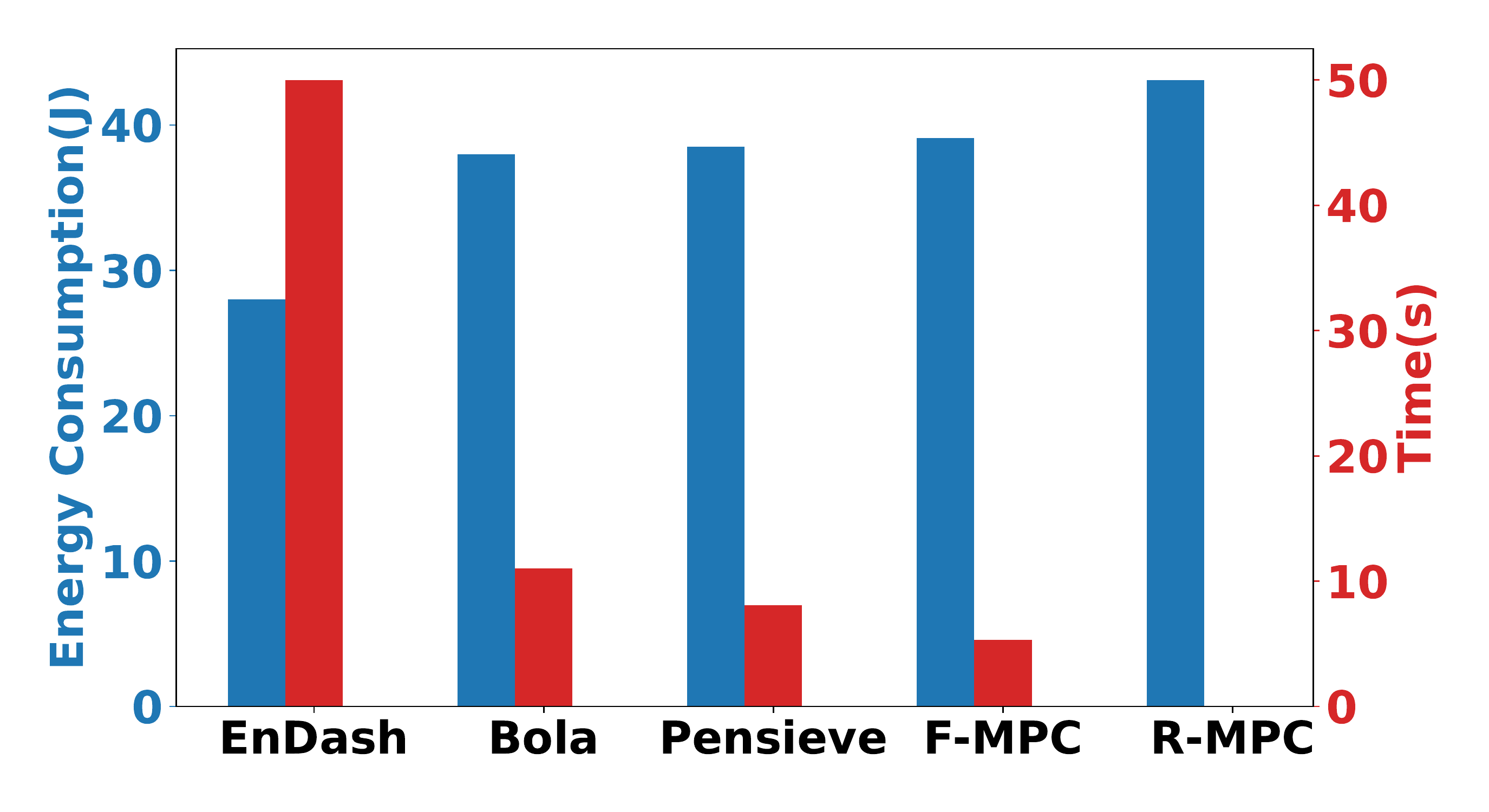}
	\caption{Energy Consumption and Extra Playtime obtained w.r.t. \ac{RMPC}, which has the highest energy consumption}\vspace*{-0.5cm}
	\label{fig:vid_time_save}
\end{figure}

\subsection{Gain from Energy Savings}
This section analyzes the gain in video playback time achieved by EnDASH-5G.  Fig.~\ref{fig:vid_time_save} shows the energy consumed by the ABR algorithms, averaged across all network loads, \ac{UE} speed, and number of buildings in the simulation area. It is seen that \ac{RMPC} consumes the highest energy, hence, we  calculate the gain in playback time with respect to  \ac{RMPC}. Let $\mathcal{E}_a$ denote the energy consumed by an ABR algorithm, where $a\in\left\{\mathtt{FMPC,\ RMPC, \ BOLA, \ and\ Pensieve}\right\}$. As playback time is inversely proportional to energy consumption, for any ABR algorithm $a$, the playback time is $T_a = \frac{k}{\mathcal{E}_a}$. We may assume that $k$ is a constant specific to a \ac{UE} handset-service provider-mobile network combination. Therefore,  for algorithm $a$, extra playback time w.r.t. \ac{RMPC} is, $T_a - T_\mathtt{RMPC}$, where $T_\mathtt{RMPC} = \frac{k}{\mathcal{E}_\mathtt{RMPC}} = \frac{T_a \times \mathcal{E}_a}{\mathcal{E}_\mathtt{RMPC}}$. 
Thus, the extra playback time w.r.t. \ac{RMPC} is given by
\begin{align}
    \label{eq:extra_playtime}
    &\text{Extra play time}  =\frac{\mathcal{E}_\mathtt{RMPC} - \mathcal{E}_a}{\mathcal{E}_\mathtt{RMPC}} \times T_a
\end{align}
The extra time gained by each algorithm over \ac{RMPC} when playing a 100-second video  is also shown in Fig.~\ref{fig:vid_time_save}.
It is observed that while EnDASH-5G gains 42 seconds, BOLA and Pensieve gain $9$ and $6$ seconds of extra playback time than \ac{RMPC}. The increase in playback time will also increase with the length of the video. Thus, one may infer that while streaming videos, EnDASH-5G can be used to improve the battery backup of 5G smartphones.
%we discuss the gain in video playback time achieved by using EnDASH. 
%FastMPC has the highest energy consumption among all algorithms.
 
  % concerning FastMPC while streaming a 2200 second video.
  
%\indent Thus, our EnDASH algorithm offers significant energy savings in conjunction with a reasonably high \ac{QoE}. The performance of EnDASH, however, depends heavily on the throughput prediction engine, various aspects of which are discussed next. 
\subsection{Performance of Throughput Prediction Engine } 
 This section evaluates the performance of the proposed transfer learning-based throughput prediction algorithm for 5G mmWave networks. The throughput prediction engine is the heart of EnDASH-5G. The results in this section have been generated for a historical window size of $H = 30s$ and a future window size of $ \mathcal{W} = 30s$. To evaluate the goodness of prediction, we use $R^2$ score and \textbf{correlation coefficient} as evaluation metrics. We have explored all the three induction bias transfer techniques as discussed in \S\ref{sec:TL_LSTM}. 
 %First, we train the LSTM, for which the architecture is outlined in \S\ref{sec:TL_LSTM}, using the `4G-India' dataset. We have then  used the same weights of the LSTM for 5G throughput prediction. This is the \textit{weight-transfer} method. We have then retrained the last LSTM layer (partial retraining) and also the entire LSTM (full retraining). 
The prediction accuracy  for the simulated 5G mmWave dataset is tabulated in  Table~\ref{tab:pred_dataset}. It is observed that the method of weight transfer performs reasonably well, with $R^2$ scores of 82.17\%. The performance improves further by 7\% by partial retraining of the layers. Complete retraining of all layers improves the $R^2$ score compared to partial retraining by 2\%. In this work, we have partially retrained the LSTM layers for 5G throughput prediction in the final implementation of the proposed system.
 
%  \begin{table}[t]
%      \centering
%      \caption{Prediction score with ns3 simulated 5G traces and  CRAWDAD Elasticmon5G dataset}
%      \label{tab:pred_dataset}
%      \begin{tabular}{|c|c|c|c|c|}
%         \hline
%          \multirow{2}{*}{Technique} &  \multicolumn{2}{|c|}{Simulated 5G} & \multicolumn{2}{|c|}{CRAWDAD 5G~\cite{crawdad-eurecom-elasticmon2019}} \\
%           & $R^2$ score &  Corr. Coeff. & $R^2$ score &  Corr. Coeff. \\ \hline\hline
%           Weight Transfer & 78.58\% & 88.38\%  & 79.34\% & 89.71\%  \\ \hline 
%           Partial Layers & 83.12\% & 93.42\% & 81.33\% & 91.65\%  \\ \hline
%           All Layers & 85.53\% & 95.75\%  & 83.17\% & 91.67\%  \\ \hline
%      \end{tabular}
%  \end{table}
 
% \usepackage{multirow}
\begin{table}[t]
     \centering
     \caption{Prediction score with ns-3 simulated 5G trace datasets}
     \label{tab:pred_dataset}
\begin{tabular}{|c|cc|cc|}
\hline
\multirow{2}{*}{Technique} & \multicolumn{2}{c|}{Simulated 5G} \\
 & \multicolumn{1}{c|}{$R^2$ score} & Pearson Corr. Coeff.  \\ \hline
Weight Transfer & \multicolumn{1}{c|}{82.17\%} & 93.53\% \\ \hline
Partial Layers & \multicolumn{1}{c|}{89.32\%} & 95.39\% \\ \hline
All Layers & \multicolumn{1}{c|}{91.27\%} & 95.91\%  \\ \hline
\end{tabular}
\end{table}
%  We observe that using the 4G model directly for prediction on 5G data has pretty good performance; we also see that fine-tuning helps improve performance even with limited 5G data and training time.
%  The Table. \ref{table:crawdad_pred} summarizes the results for the CRAWDAD datasets.
 
%  The results on this datasets are similar to the previous case however the improvement in performance using fine tune is not as significant as in the previous case.
\begin{figure}[!ht]
    \centering
    \includegraphics[width=0.6\linewidth]{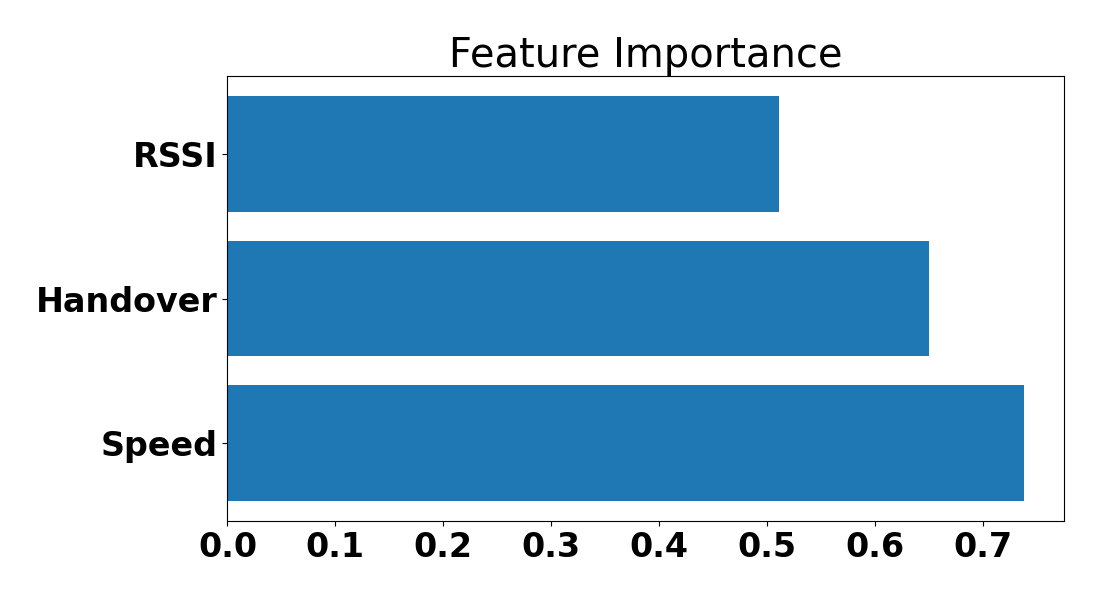}
    \caption{Feature importance in SHAP (SHAPley Additive Explanations) values of the input parameters for deciding throughput; Number of Handovers, signal strength(RSSI), and speed of the User Equipment}
    %\vspace*{-0.5cm}
    \label{fig:feature_imp}
\end{figure}

Fig.~\ref{fig:feature_imp} shows the feature importance, in SHAP (SHAPley Additive Explanations) values~\cite{lundberg2017unified}, of different input parameters, such as the \ac{UE} speed, number of handovers, and RSSI when predicting the throughput.  It may be noted that speed has the highest importance as a feature.  Vehicular speed directly impacts the link condition of the channel, which in turn affects the throughput.
%The reason for this may be attributed to the high values of speed taken for simulation purposes. 
This points to the absolute necessity of taking mobility into account for predicting throughput.

	\section{Conclusion}\label{section:conclusion}
%In many developing countries, 4G coverage is not ubiquitous. As a result, users often have to fall back to legacy networks that support lower throughput, resulting in higher energy consumption and lower QoE. 
This paper proposed EnDASH-5G -- an energy-aware ABR video streaming algorithm that minimizes energy consumption while not compromising the QoE of users under mobility in 5G mmWave networks. It exploits the high throughput regions in the user's trajectory for aggressive fetching of video chunks, thereby reducing energy consumption even in regions with the limited network coverage. To achieve this, it intelligently tunes the playback buffer length to the average predicted network throughput and then resorts to optimal bitrate selection for video chunks. As a result, the buffer length no longer remains fixed. The network throughput is predicted by EnDASH-5G  using a novel transfer learning approach.  
EnDASH-5G offers an additional 36 seconds of playback buffer time when playing a 100-second video, compared to the popular Pensieve algorithm, albeit with a marginal reduction in QoE. EnDASH-5G being tunable, can be designed to adapt to a specific requirement, such as energy or QoE maximization. This, in addition with a real-life implementation of EnDASH-5G and investigating the corresponding improvement in energy efficiency, will be our immediate future work. 
	
	\bibliographystyle{IEEEtran}
	\bibliography{refs}
	
	%\newpage

\vspace{-1cm}
\begin{IEEEbiography}[{\includegraphics[width=1in]{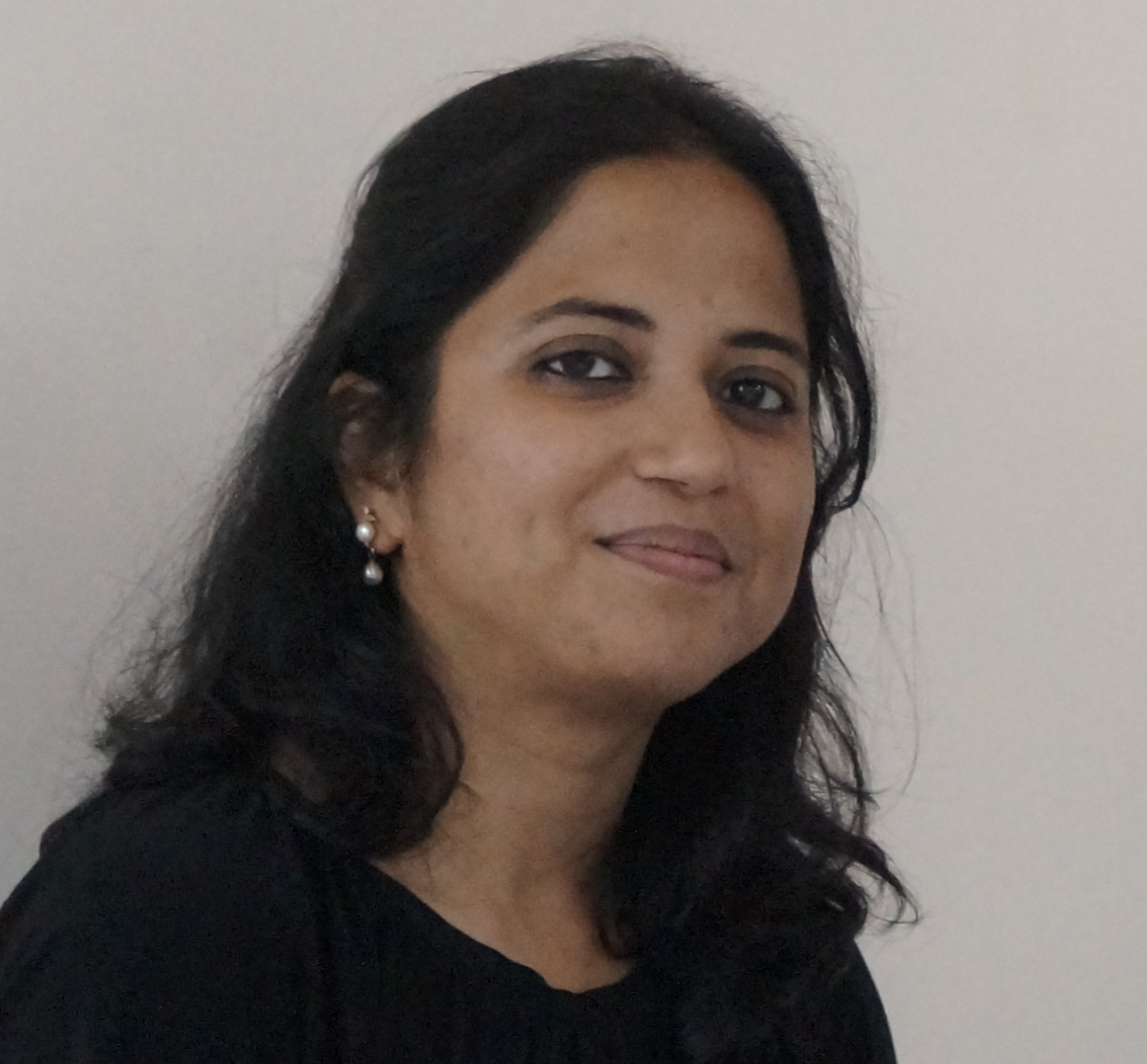}}]
{Dr. Basabdatta Palit}
    is currently a faculty in the Department of Electronics and Telecommunication Engineering at IIEST, Shibpur, India. She has received her Ph.D. from IIT Kharagpur, India in 2018, her M.E. from Jadavpur University,  India, in 2011, and her her B.Tech from West Bengal University of Technology in 2009.  Her research interests include Scheduling and Resource Allocation, Vehicular Communication, Energy Efficiency, URLLC.
\end{IEEEbiography}
 %\vspace{-1cm}

\begin{IEEEbiography}[{\includegraphics[width=1in]{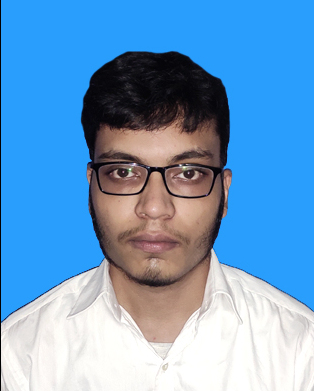}}]
{Argha Sen}
received his B.Tech in Electronics and Communication Engineering from the National Institute of Technology Durgapur, India, in 2020. He is currently pursuing his Ph.D. in the Department of Computer Science and Engineering at the IIT Kharagpur, India. His research interests include the Next-generation cellular networks, Energy Efficient Core Network Design, and ABR Streaming.
\end{IEEEbiography} 
 % \vspace{-1cm}
 
\begin{IEEEbiography}[{\includegraphics[width=1in]{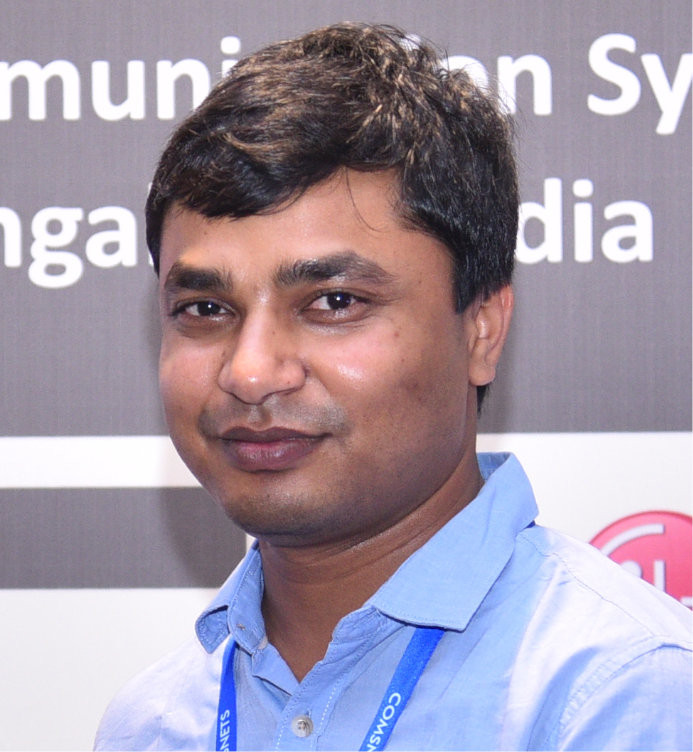}}]
{Dr. Abhijit Mondal}
received his Ph.D. degree from IIT Kharagpur in 2021. He completed his B.Tech from Maulana Abul Kalam Azad University Of Technology, West Bengal, India in 2010 and his M.Tech from IIT Guwahati, India in 2012. He is currently working as a Senior Engineer at VDX.tv. Previously, he has worked in Rovi (now TiVo) as a Software Engineer.
\end{IEEEbiography} 
% \vspace{-1cm}
  
\begin{IEEEbiography}[{\includegraphics[width=1in]{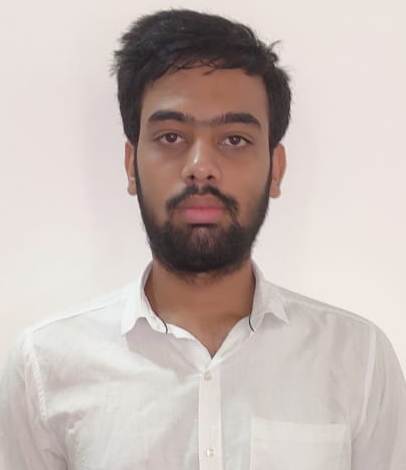}}]
{Ayan Zunaid}
has received his B.Tech and M.Tech in Computer Science and Engineering from IIT, Kharagpur,  India, in 2021. He is currently working with Flipkart as a software development engineer 1. His research interests include Machine Learning and Big Data Analytics.
\end{IEEEbiography}   
%  \vspace{-1cm}
  
\begin{IEEEbiography}[{\includegraphics[width=1in]{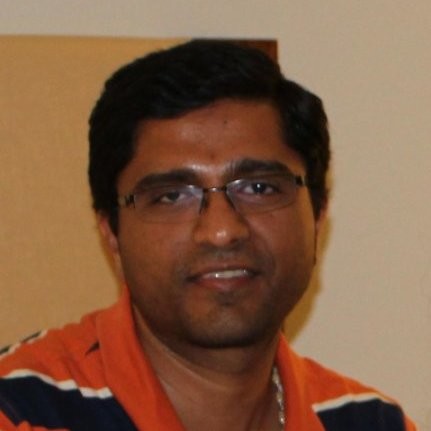}}]
{Jay Jayatheerthan}
 has received his B.Tech from P.S.G College of Technology, Coimbatore, India and Masters in Engineering Management from UT, Austin. He is currently a technologist with Intel India, managing a team working in 5G, Telecom and Networking industry.  His research interest include optimizing 5G networks for energy efficiency and using Intel Xeon SoCs to newer workloads and market segments.
\end{IEEEbiography}  
% \vspace{-1cm}

\begin{IEEEbiography}[{ \includegraphics[width=1in]{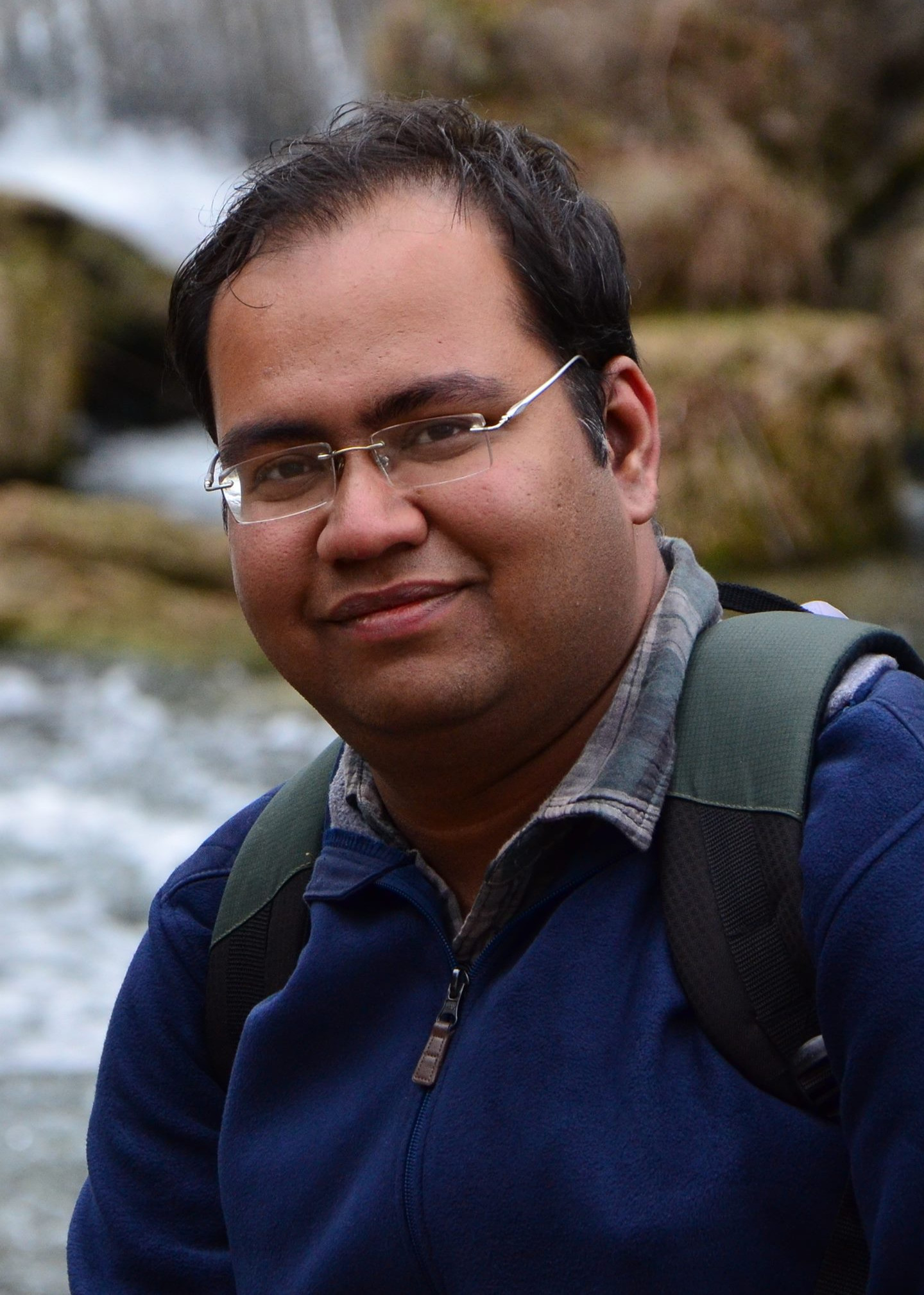}}]%
{Dr. Sandip Chakraborty}
 is working as an Associate Professor in the Department of Computer Science and Engineering at IIT Kharagpur. He obtained his B.E from Jadavpur University in 2009, and M.Tech and PhD degree both from IIT Guwahati, India, in 2011 and 2014, respectively. His primary research interest  is in the intersection of Computer Systems, Pervasive Computing, and Human-Computer Interaction. He is one of the founding members of ACM IMOBILE, the ACM SIGMOBILE chapter in India. He is working as an Area Editor of Elsevier Ad Hoc Networks and Elsevier Pervasive and Mobile Computing journal. 
\end{IEEEbiography}

% \vfill
	
\end{document}